\documentclass[manuscript]{aastex62}
\usepackage[encapsulated]{CJK}
\usepackage{mathrsfs}
\usepackage{subfigure}
\usepackage{amsmath}
\usepackage{longtable}
\usepackage{booktabs}
\usepackage{threeparttable}
\usepackage{hyperref} 
\usepackage{overpic}
\def\etal {et al.~}

\newbox\grsign \setbox\grsign=\hbox{$>$} \newdimen\grdimen \grdimen=\ht\grsign
\newbox\laxbox \newbox\gaxbox
\setbox\gaxbox=\hbox{\raise.5ex\hbox{$>$}\llap
     {\lower.5ex\hbox{$\sim$}}}\ht1=\grdimen\dp1=0pt
\setbox\laxbox=\hbox{\raise.5ex\hbox{$<$}\llap
     {\lower.5ex\hbox{$\sim$}}}\ht2=\grdimen\dp2=0pt

\shorttitle{HINSA}
\shortauthors{Li \etal}

\definecolor{malachite}{rgb}{0.34, 0.7, 0.22}

\begin{document}
\begin{CJK*}{UTF8}{gbsn}

\title{H~\textsc{i} Narrow-Line Self-Absorptions Toward the High-Mass Star-Forming Region G176.51+00.20}

\correspondingauthor{Yingjie Li}
\email{liyj@pmo.ac.cn, xuye@pmo.ac.cn}

\author{Yingjie Li}\affiliation{Purple Mountain Observatory, Chinese Academy of Sciences, Nanjing 210008, China}

\author{Ye Xu}
\affiliation{Purple Mountain Observatory, Chinese Academy of Sciences, Nanjing 210008, China}

\author{Jin-Long Xu}
\affiliation{National Astronomical Observatories, Chinese Academy of Sciences, Beijing 100101, China}
\affiliation{CAS Key Laboratory of FAST, National Astronomical Observatories, Chinese Academy of Sciences, Beijing 100101, China}

\author{Dejian Liu}
\affiliation{Purple Mountain Observatory, Chinese Academy of Sciences, Nanjing 210008, China}
\affiliation{University of Science and Technology of China, Hefei, Anhui 230026, China}

\author{Jingjing Li}
\affiliation{Purple Mountain Observatory, Chinese Academy of Sciences, Nanjing 210008, China}

\author{Zehao Lin}
\affiliation{Purple Mountain Observatory, Chinese Academy of Sciences, Nanjing 210008, China}
\affiliation{University of Science and Technology of China, Chinese Academy of Sciences, Hefei, Anhui 230026, China}

\author{Peng Jiang}
\affiliation{National Astronomical Observatories, Chinese Academy of Sciences, Beijing 100101, China}
\affiliation{CAS Key Laboratory of FAST, National Astronomical Observatories, Chinese Academy of Sciences, Beijing 100101, China}

\author{Shuaibo Bian}
\affiliation{Purple Mountain Observatory, Chinese Academy of Sciences, Nanjing 210008, China}
\affiliation{University of Science and Technology of China, Hefei, Anhui 230026, China}

\author{Chaojie Hao}
\affiliation{Purple Mountain Observatory, Chinese Academy of Sciences, Nanjing 210008, China}
\affiliation{University of Science and Technology of China, Hefei, Anhui 230026, China}

\author{Xiuhui Chen}
\affiliation{College of Mathematics and Physics, Hunan University of Arts and Science, Changde, Hunan 415300, China}

\begin{abstract}
Using the Five-hundred-meter Aperture Spherical radio Telescope (FAST) 19-beam tracking observational mode, high sensitivity and high-velocity resolution H~\textsc{i} spectral lines have been observed toward the high-mass star-forming region G176.51+00.20. This is a pilot study of searching for H~\textsc{i} narrow-line self-absorption (HINSA) toward high-mass star-forming regions where bipolar molecular outflows have been detected. This work is confined to the central seven beams of FAST. Two HINSA components are detected in all seven beams, which correspond to a strong CO emission region (SCER; with a velocity of $\sim$ $-$18 km s$^{-1}$) and a weak CO emission region (WCER; with a velocity of $\sim$ $-$3 km s$^{-1}$). The SCER detected in Beam 3 is probably more suitably classified as a WCER.
In the SCER, the HINSA is probably associated with the molecular material traced by the CO. The fractional abundance of HINSA ranges from  $\sim 1.1 \times 10^{-3}$ to $\sim 2.6 \times 10^{-2}$. Moreover, the abundance of HINSA in Beam 1 is lower than that in the surrounding beams (i.e., Beams 2 and 4--7). This possible ring could be caused by ionization of H~\textsc{i} or relatively rapid conversion from H~\textsc{i} to H$_2$ in the higher-density inner region. In the WCER (including Beam 3 in the SCER), the HINSA is probably not associated with CO clouds, but with CO-dark or CO-faint gas. 
\end{abstract}



\section{Introduction}\label{sec-introduction}

Stars form in molecular clouds, and it is important for star formation to synthetize molecular H$_2$ from atomic H. However, 
it is difficult to directly extract individual components from H~\textsc{i} emission and correlate them with molecular clouds \citep{Li-Goldsmith2003, Liu+2022}. In such conditions, H~\textsc{i} self-absorption (HISA) lines have become a very useful tracer of H \textsc{i}. HISA occurs in environments where cold H~\textsc{i} gas is in the front of a warmer emission background \citep[e.g.][]{Knapp1974}; such conditions are ubiquitous in the Milky Way \citep[e.g.,][]{Heeschen+1955, Gibson+2000, Wang+2020}. HISA is associated with both CO-dark clouds and molecular clouds with strong CO emission \citep[e.g.,][]{Hasegawa+1983, Gibson+2000, Gibson2010}. H~\textsc{i} narrow-line self-absorption (HINSA), a special case of HISA, is usually associated with CO clouds, and its
linewidth is comparable or smaller than that of CO \citep[see][]{Li-Goldsmith2003, Goldsmith-Li2005, Tang+2016, Wang+2020}. HINSA has been found to be an excellent tracer of molecular clouds \citep{Li-Goldsmith2003}, and this tight correlation has been confirmed by numerous studies \citep[e.g.,][]{Krco-Goldsmith2010, Tang+2020, Liu+2022}.

Numerous surveys have provided abundant knowledge about the environment and physical properties of the regions producing HINSA. For instance, the detection rate of HINSA is $\sim$ 77\% \citep{Li-Goldsmith2003} in optically selected dark clouds \citep{Lee-Myers1999}, $\sim$ 80\% \citep{Krco-Goldsmith2010} in molecular cores \citep[i.e., the Lynds dark cloud, which has a peak optical extinction of 6 mag
or more; see the catalog of dark clouds in][]{Lynds1962}, and $\sim$ 58\% \citep{Tang+2020} to $\sim$ 92\% \citep{Liu+2022} in Planck Galactic cold clumps \citep[PGCCs; cold and quiescent clumps in very early evolution stages of star formation; see][]{Planck-Collaboration+2011, Wu+2012, Planck-Collaboration+2016}. These results indicate that HINSA is associated with molecular clouds, especially molecular cores and/or clumps. These papers also constrained the abundance of HINSA, which ranges from $\sim$ 10$^{-4}$ to $\sim$ 10$^{-2}$ \citep[e.g.,][]{Krco-Goldsmith2010, Tang+2020, Liu+2022}; the relationship of the central velocity, linewidth, and column density between HINSA and the molecular gas traced by $^{12}$CO, $^{13}$CO, OH, etc. \citep[e.g.,][]{Li-Goldsmith2003, Tang+2020, Liu+2022} and the atomic gas traced by CI \citep[e.g.,][]{Li-Goldsmith2003}; and even a ring of enhanced HINSA abundance inside of a dark molecular cloud \citep[i.e., in B227; see][]{Zuo+2018}. 

Star-forming regions characterized by molecular outflows have also been incorporated into samples arising from the search for HINSA \citep[e.g.,][]{Liu+2022}. In this study, the corresponding detection rate of HINSA was only 25\% (i.e., one in four sources characterized by molecular outflows). As such, one of our goals is to enlarge the sample of known star-forming regions characterized by molecular outflows. A good choice is the nine regions where outflowing gases traced by $^{12}$CO, $^{13}$CO, HCO$^+$, and CS have been recently detected with high sensitivity (i.e., main beam rms noise
of tens of mK) with the 13.7 m millimeter telescope of the Purple Mountain Observatory in Delingha by \citet{Liu+2021}. In one of the nine regions, i.e., G176.51+00.20, a neutral stellar wind traced by atomic hydrogen (H~\textsc{i} wind) has also been detected (Y. J., Li et al., 2022, ApJ, accepted). Therefore, studying HINSA toward G176.51+00.20 not only would be a bridge to connect the atomic and molecular gas therein, but may also be helpful for studying the relationship between H~\textsc{i} winds and molecular outflows.

G176.51+00.20, an active high-mass star-forming region, is located 1.8 kpc from Earth \citep{Moffat+1979, Snell+1988}. The dense NH$_3$ core in the center of this region has become synonymous with this region \citep[e.g.,][]{Torrelles+1992, Chen+2003, Jiang+2013}. The central engine of this massive star-forming region was identified as a zero-age main-sequence B3 star by using a 3.6 cm map produced from Very Large Array observations \citep{Torrelles+1992b}. A more detailed description of this region was provided by \citet{Dewangan2019}, where the center of the region of study in this work was marked as ``H~\textsc{ii} region'' in their figure 1(a).

The remainder of this paper is organized as follows. In Section \ref{sec-data}, we describe the data used in this work. Section \ref{sec-HINSA} presents the results of the HINSA survey toward G176.51+00.20, and the physical properties of the HINSA features and those of the corresponding molecular lines. In Section \ref{sec-discuss}, we discuss the association between the HINSA features and the molecular clouds, and estimate the abundance of the HINSA features, and summarize our results.

\section{Data}\label{sec-data}

\subsection{H~\textsc{i} Observations with FAST} \label{sec-data-HI}

The most sensitive ground-based single-dish radio telescope is the Five-hundred-meter Aperture Spherical radio Telescope\footnote{\url{https://fast.bao.ac.cn/}}, which located in Guizhou Province of southwest China \citep{Nan2006, Nan+2011, Qian+2020}. By using the 19-beam receiver equipped on FAST \citep[see][]{Lidi+2018, Jiang+2019, Jiang+2020}, we observed highly sensitive and high-velocity resolution H~\textsc{i} spectra toward G176.51+00.20. The mean main beam rms noise for the 19 obtained spectra is $\sim$ 7 mK @ 0.1 km s$^{-1}$ resolution. The observations were conducted on August 19th and 20th, 2021, with total integration time is 335 minutes. The half-power beam width (HPBW) is $\sim$ 2.9$'$ at 1.4 GHz, and the pointing error is $\sim$ 0.2$'$ \citep[see][]{Jiang+2019, Jiang+2020}. 

\subsection{Summary of Molecular Lines Observed Toward G176.51+00.20}\label{sec-data-molecular}

Five high-sensitivity molecular lines were observed toward G176.51+00.20 (with a cell of $\sim$ $15' \times 15'$) by using the 13.7 m telescope: $^{12}$CO ($J = 1-0$)(115.271 GHz), $^{13}$CO ($J = 1-0$)(110.201 GHz), and C$^{18}$O ($J = 1-0$)(109.782 GHz) observed from 2019 July to November; HCO$^+$ ($J = 1-0$) (89.189 GHz) observed from 2019 November to 2020 February; and CS ($J = 2-1$)(97.981 GHz) observed in 2020 May \citep[for more details, see tables 2--3 of][]{Liu+2021}. The velocity resolution of these lines ranges from 0.159 to 0.212 km s$^{-1}$ (the corresponding channel width is 61 kHz), and the HPBW ranges from 49$''$ to 61$''$. The lines were observed with the on-the-fly mode, and gridded to 30$''$ pixels. The rms noises range from $\sim$ 60.9 to $\sim$ 15.3 mK \citep[for more details see][]{Liu+2021}.

\section{Data Analysis and Results}\label{sec-HINSA}

Because the molecular line observations are confined to the central $\sim$ $15' \times 15'$ region centered at (R.A., Decl.) (J2000) = (05$^{\mathrm{h}}$37$^{\mathrm{m}}$52$^{\mathrm{s}}$.1, 32$^{\circ}$00$'$03$''$.0), the HINSA survey is also limited to the corresponding regions; i.e., central seven beams (Beams 1--7; see Figure \ref{fig-12co-moment}(a)). We followed the method of \citet{Liu+2022} to extract HINSA features. To extract HINSA, it was required to understand the physical properties of the emission regions, at least the central velocity, $v_{\mathrm{c,m}}$, velocity dispersion, $\sigma_{\mathrm{m}}$, and excitation temperature, $T_{\mathrm{ex}}$ \citep[i.e., the initial conditions to extract HINSA; see][]{Liu+2022}. These three physical parameters were obtained by using the high-sensitivity molecular line observations of the high-mass star-forming region G176.51+00.20 made by \citet{Liu+2021}.

\subsection{Overview of the Region with High-Sensitivity Molecular-Line Observations}\label{sec-HINSA-molecular-environment}

From the study of molecular outflows toward G176.51+00.20 by \citet{Liu+2021}, the velocity of the most conspicuous components is $\sim$ $-$18 km s$^{-1}$, which corresponds to a high-mass star-forming region with a strong bipolar outflow \citep[see][]{Snell+1988, Zhang+2005, Liu+2021}. In fact, there is also a weak component with a velocity of $\sim$ $-3$ km s$^{-1}$ in the region close to the edge of the observed region. Throughout this work, we have divided the observed region into two parts: a strong CO emission region (hereafter SCER) with a velocity of $\sim$ $-$18 km s$^{-1}$ and a weak CO emission region (WCER) with a velocity of $\sim$ $-$3 km s$^{-1}$.

\begin{figure*}[!htb]
	\centering
\subfigure[WCER]{\includegraphics[height=0.37\textwidth]{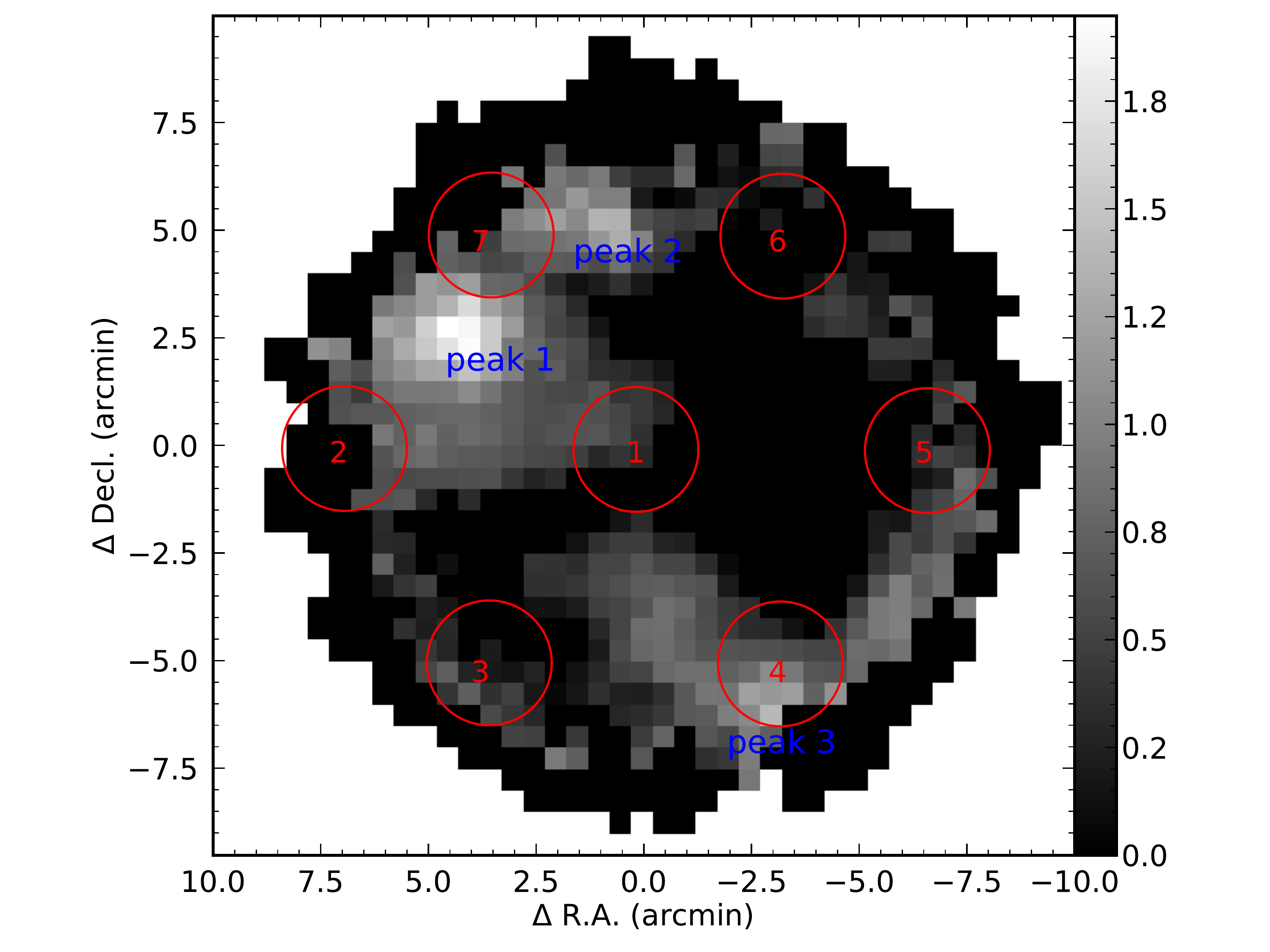}}
\subfigure[SCER]{\includegraphics[height=0.37\textwidth]{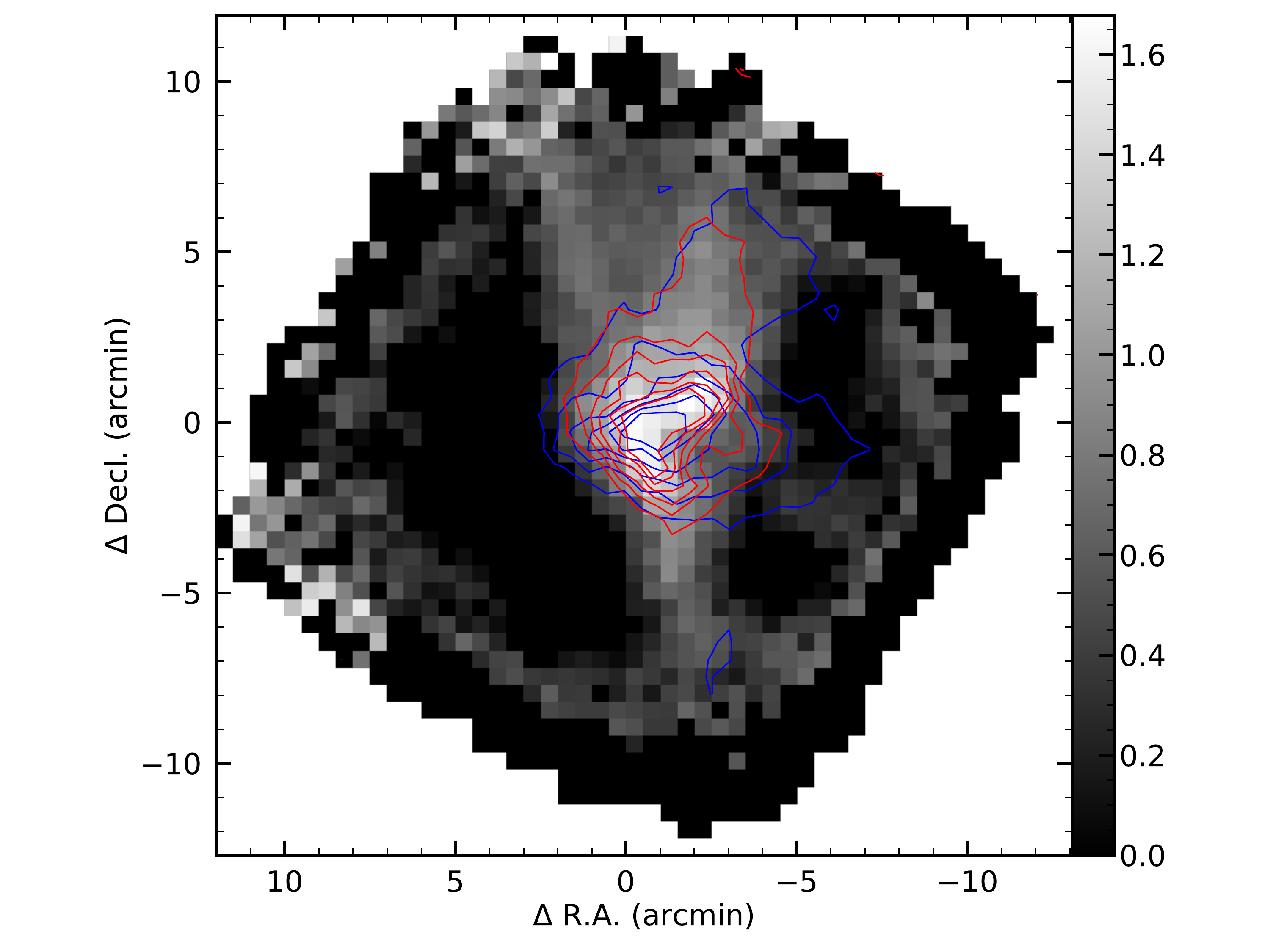}}
\caption{Integrated intensity map of the WCER (left) and SCER (right). Left: the background map is the integrated intensity map of $^{12}$CO in the range of [$-5$, $-1$] km s$^{-1}$. Peaks 1--3 label the peaks in the WCER (i.e., corresponding to the intensity peaks in the $^{12}$CO map), and the red circles and numbers present the positions and indices of the central seven beams of FAST. Right: the background is an integrated intensity map of C$^{18}$O in the range of [$-30$, $-10$] km s$^{-1}$, where the blue and red contours show the integrated intensity map of $^{12}$CO and $^{13}$CO in the same velocity range as the C$^{18}$O map, respectively. Their levels are 0.3, 0.4, 0.5, 0.6, 0.7, and 0.8 times the corresponding peak intensity. The color bars in the left and right panels are in units of $\mathrm{(K\;km\;s^{-1})^{1/2}}$.}
\label{fig-12co-moment}
\end{figure*}

Figure \ref{fig-12co-moment} presents an integrated intensity map of the WCER (left) and the SCER (right). In the WCER, the $^{13}$CO emission is weak and no C$^{18}$O emission is detected. Therefore, only the integrated intensity map of $^{12}$CO, with a velocity range of [$-$5, $-$1] km s$^{-1}$, is presented here. There are three peaks in the CO map of the WCER; i.e., Peaks 1--3 (see the labels in Figure \ref{fig-12co-moment}(a)). In the SCER, there is strong $^{12}$CO (blue contours), $^{13}$CO (red contours), and C$^{18}$O (background map) emission, where the velocity range of the maps is [$-$30, $-$10] km s$^{-1}$. Because the various emission components are all connected, and it is difficult to distinguish each of them clearly, we marked the position centered at the peak intensity of C$^{18}$O as Peak 4. We note that this large emission region covers, in fact, almost the entire observed region (including the regions covered by all seven FAST beams; see below). 

In the WCER, because the line emission of $^{13}$CO is weak, and lines from other molecules (i.e., C$^{18}$O, HCO$^+$ and CS) are not detected, we only conducted Gaussian fits for the $^{12}$CO spectra. The values of $v_\mathrm{{c,m}}$ and $\sigma_{\mathrm{m}}$ are given based on those Gaussian fits. The peak brightness temperatures of the $^{12}$CO spectra, $T^{\mathrm{^{12}CO}}_{\mathrm{peak}}$, were used to calculate the value of $T_{\mathrm{ex}}$ \citep[e.g.,][]{Garden+1991} as
\begin{equation}\label{equ:tex} 
	T_{\mathrm{ex}} = \frac{5.53}{\displaystyle \ln \left(1 +  \frac{5.53}{T^{\mathrm{^{12}CO}}_{\mathrm{peak} }+ 0.82}\right)}. 
\end{equation}
The results of $v_\mathrm{{c,m}}$, $\sigma_{\mathrm{m}}$, and $T_{\mathrm{ex}}$, with 1$\sigma$ errors, are $-$3.92 $\pm$ 0.01, 0.35 $\pm$ 0.01 km s$^{-1}$, and 7.18 $\pm$ 0.18 K for Peak 1, and $-$2.20 $\pm$ 0.01, 0.44 $\pm$ 0.01 km s$^{-1}$, and 4.76 $\pm$ 0.16 K for Peak 2, and $-$2.54 $\pm$ 0.04, 0.42 $\pm$ 0.04 km s$^{-1}$, and 5.08 $\pm$ 0.31 K for Peak 3, respectively. In the SCER, the value of $T_{\mathrm{ex}}$ was also calculated based on $T^{\mathrm{^{12}CO}}_{\mathrm{peak}}$. The values of $v_\mathrm{{c,m}}$ and $\sigma_{\mathrm{m}}$ are based on the Gaussian fit of the $^{13}$CO spectrum. The values of $v_\mathrm{{c,m}}$, $\sigma_{\mathrm{m}}$, and $T_{\mathrm{ex}}$ are $-$18.53 $\pm$ 0.01, 1.13 $\pm$ 0.01 km s$^{-1}$ and 20.30 $\pm$ 0.08 K, respectively. These values were used to extract the HINSA features. 

\subsection{HINSA Features}\label{sec-HINSA-features}

To extract as many HINSA features as possible, we assumed that, for the spectrum of each FAST beam, HINSA features corresponding to each of the four components of the molecular clouds (i.e., Peaks 1--4) exist. The specific description of the method used to extract the HINSA feature is presented in Section \ref{sec-HINSA-extract}.

\begin{figure*}[!htb]
	\centering
	\subfigure[Beam 1]{\includegraphics[height=0.22\textwidth]{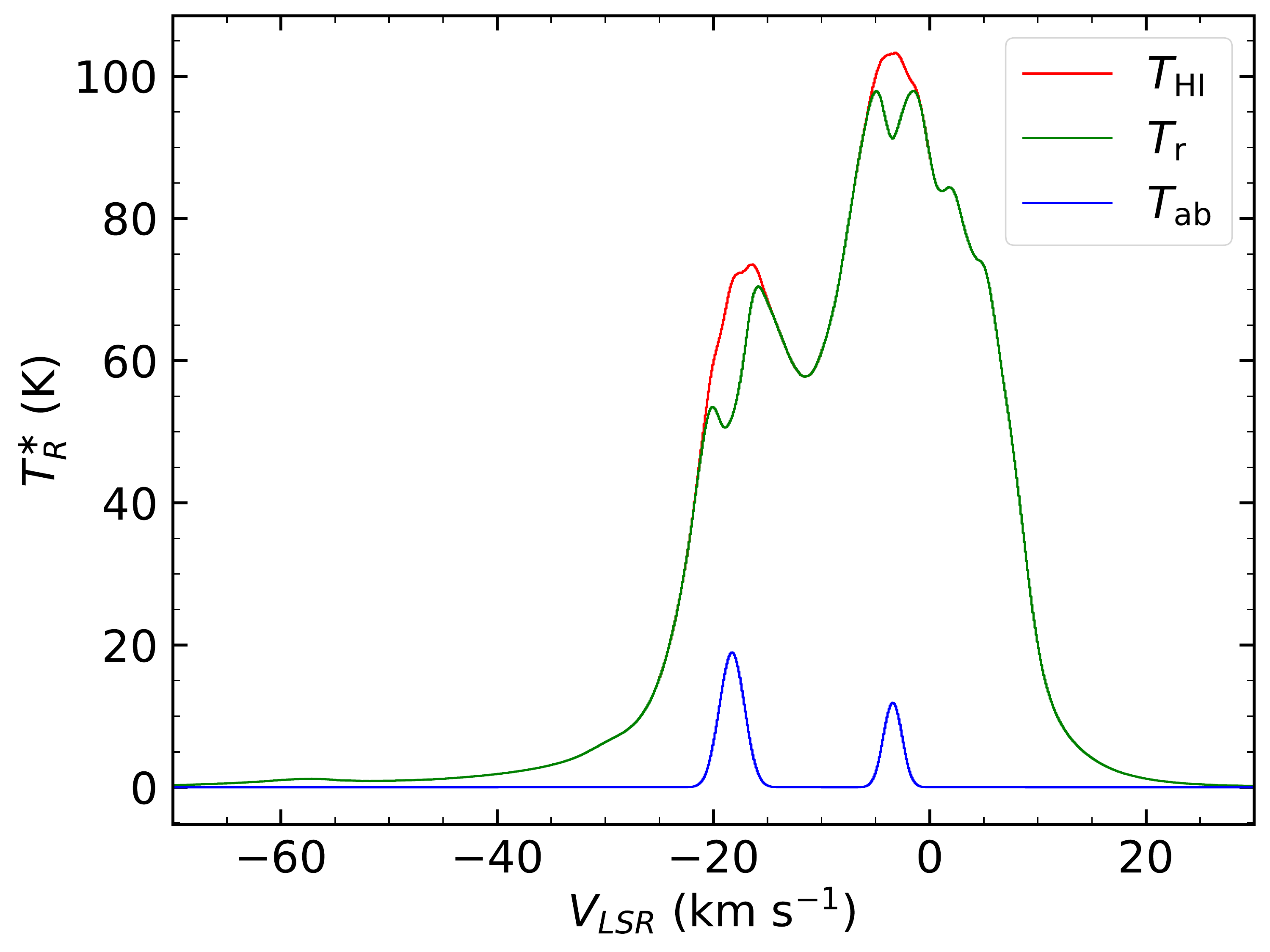}}
	\subfigure[Beam 2]{\includegraphics[height=0.22\textwidth]{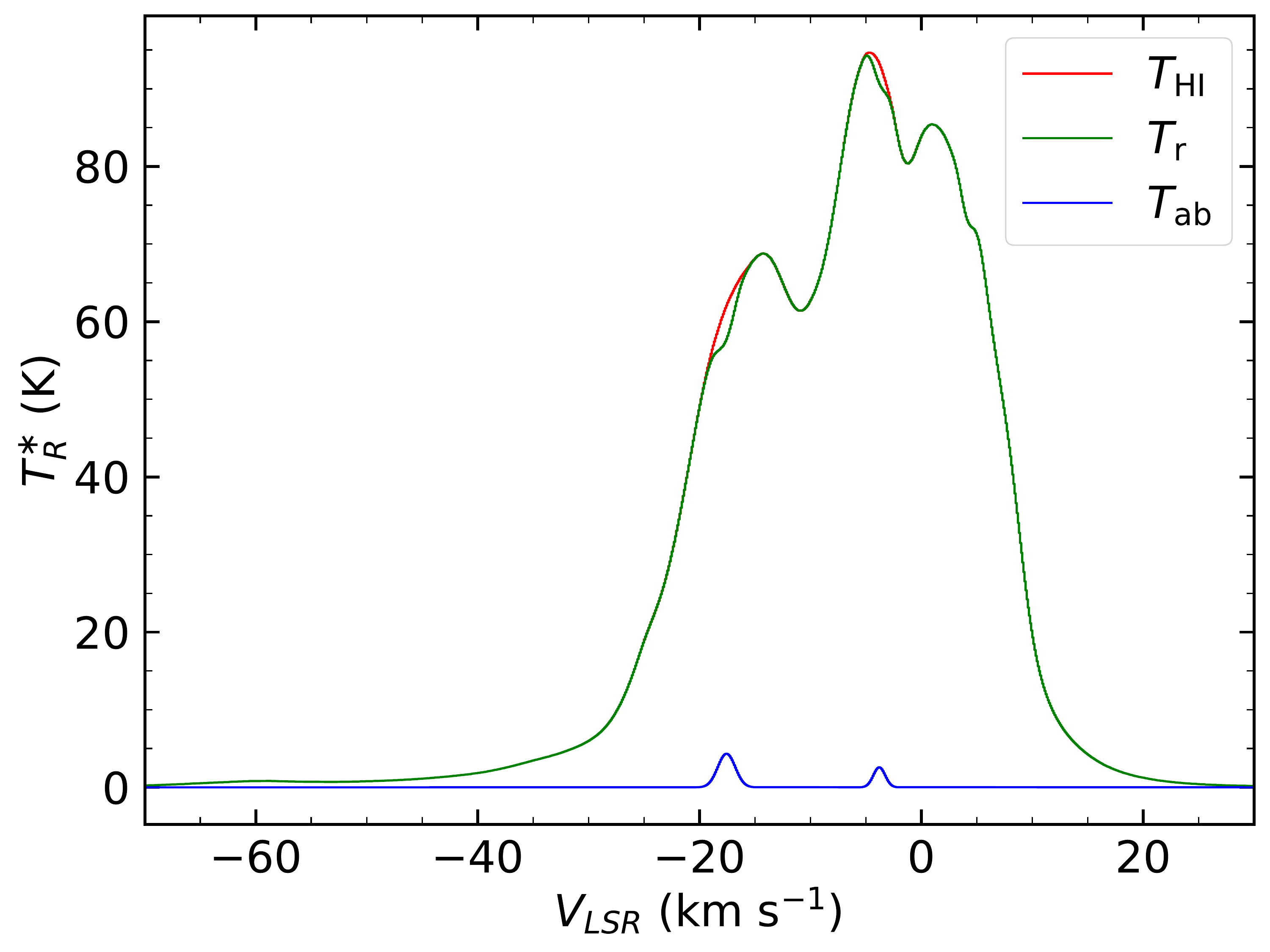}}
	\subfigure[Beam 3]{\includegraphics[height=0.22\textwidth]{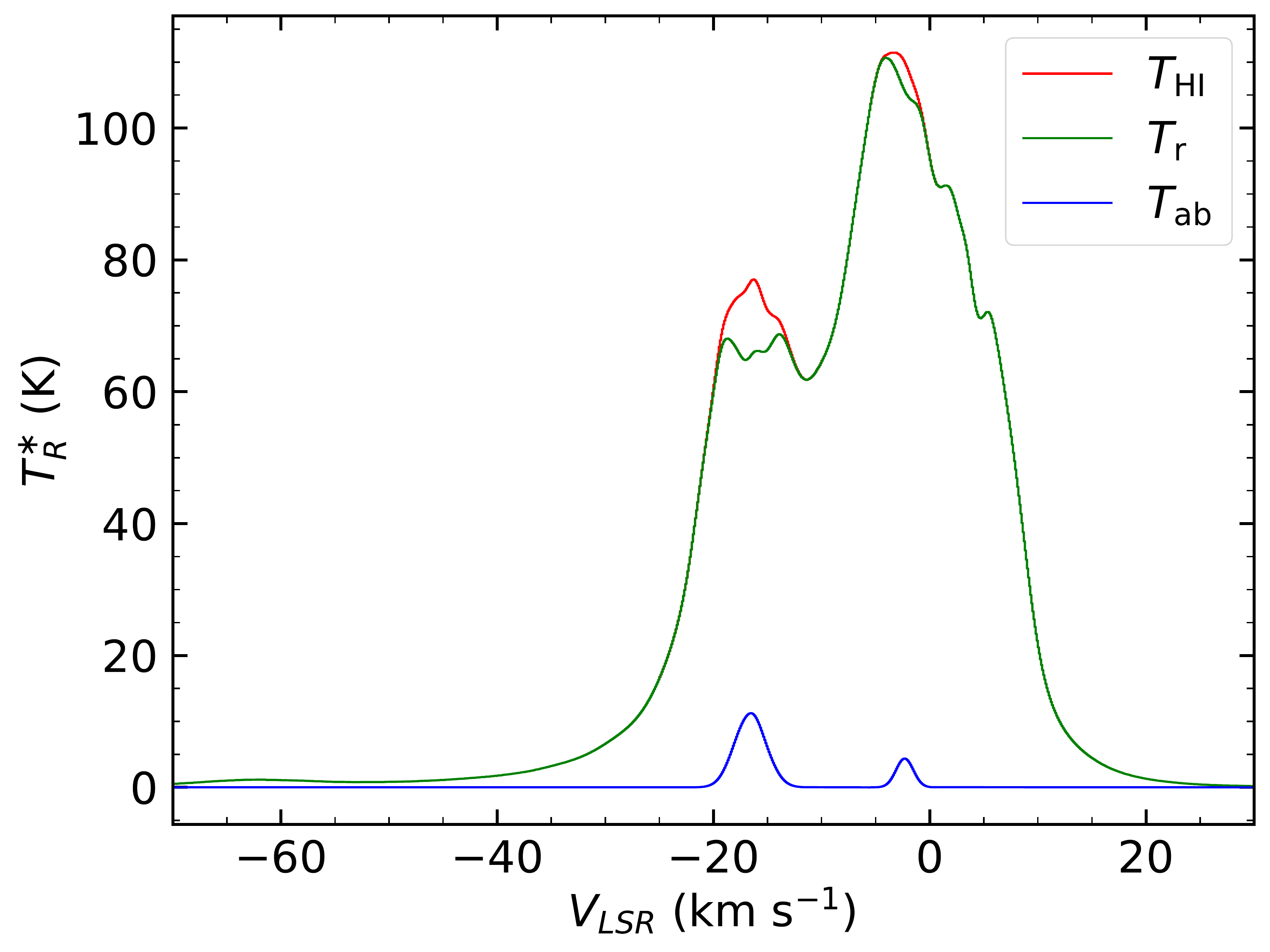}}
	\subfigure[Beam 4]{\includegraphics[height=0.22\textwidth]{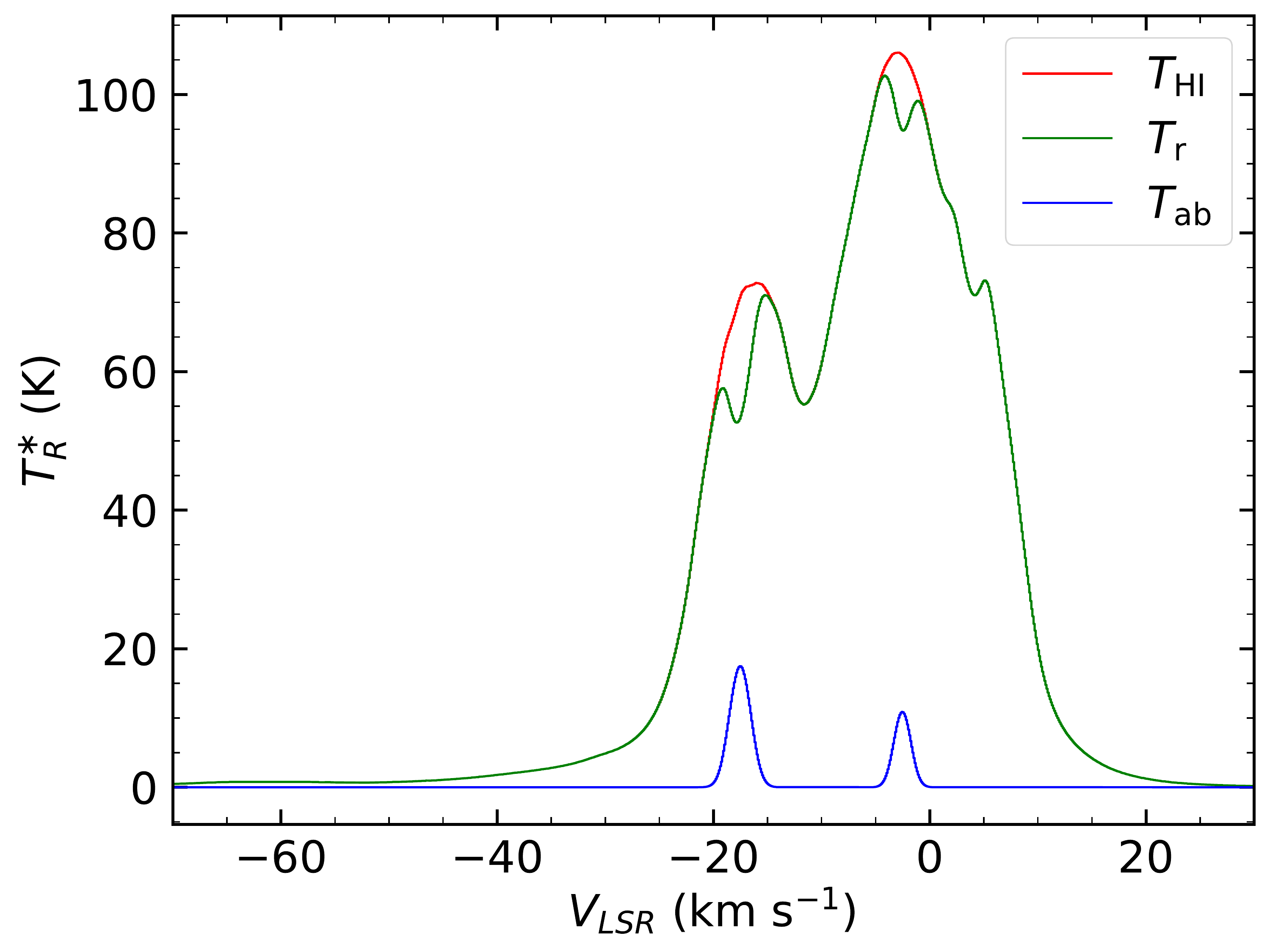}}
	\subfigure[Beam 5]{\includegraphics[height=0.22\textwidth]{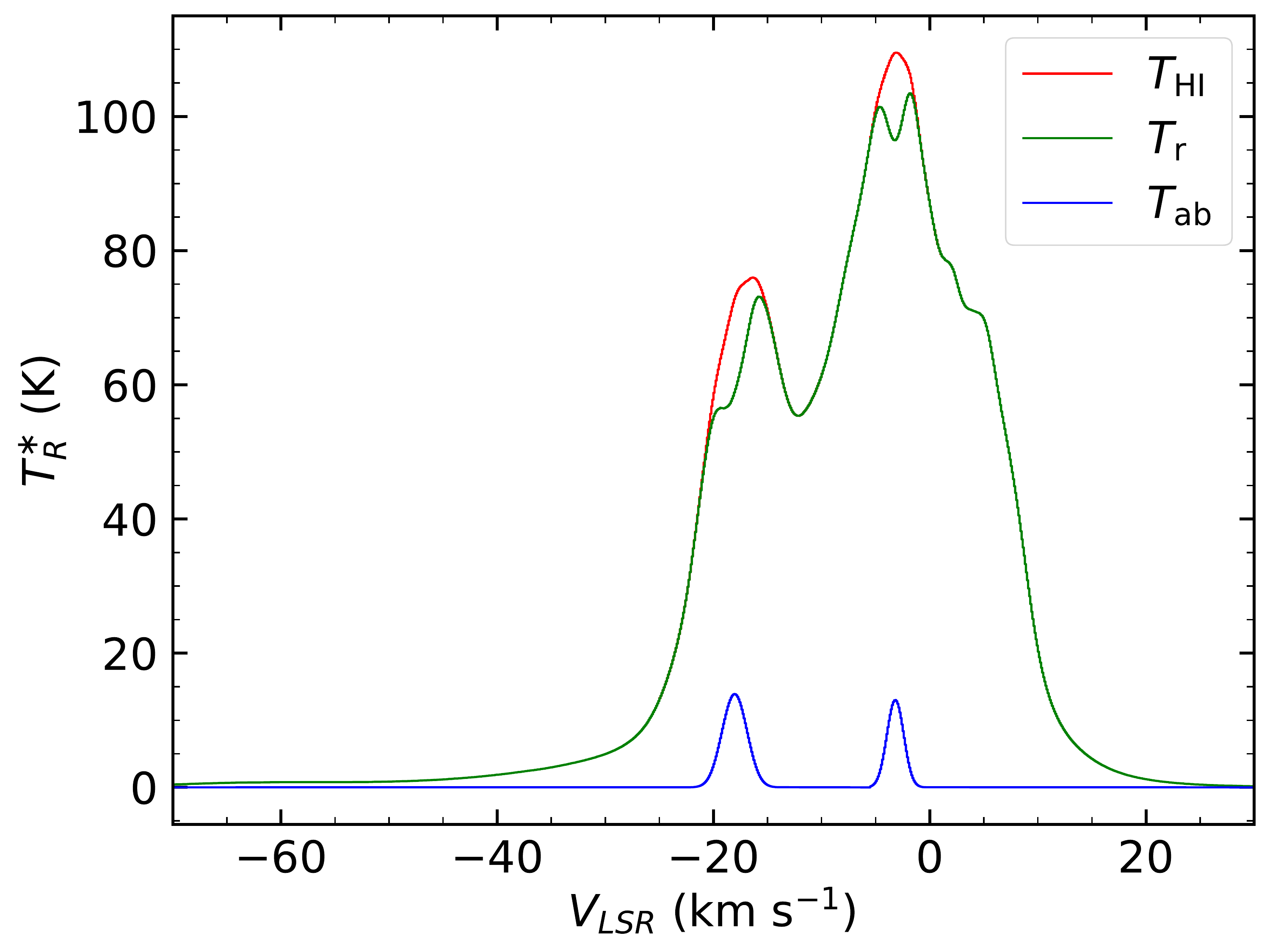}}
	\subfigure[Beam 6]{\includegraphics[height=0.22\textwidth]{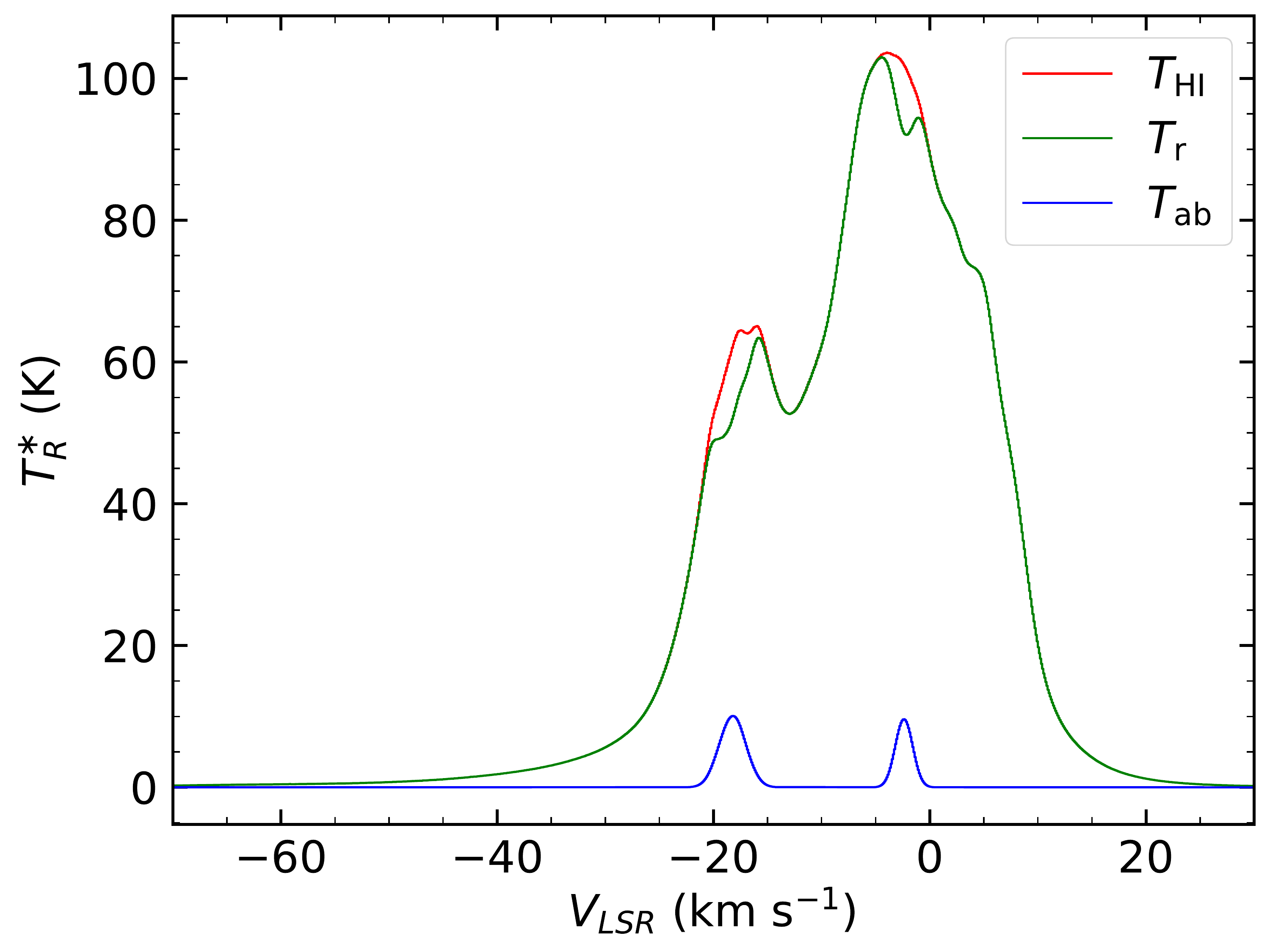}}
	\subfigure[Beam 7]{\includegraphics[height=0.22\textwidth]{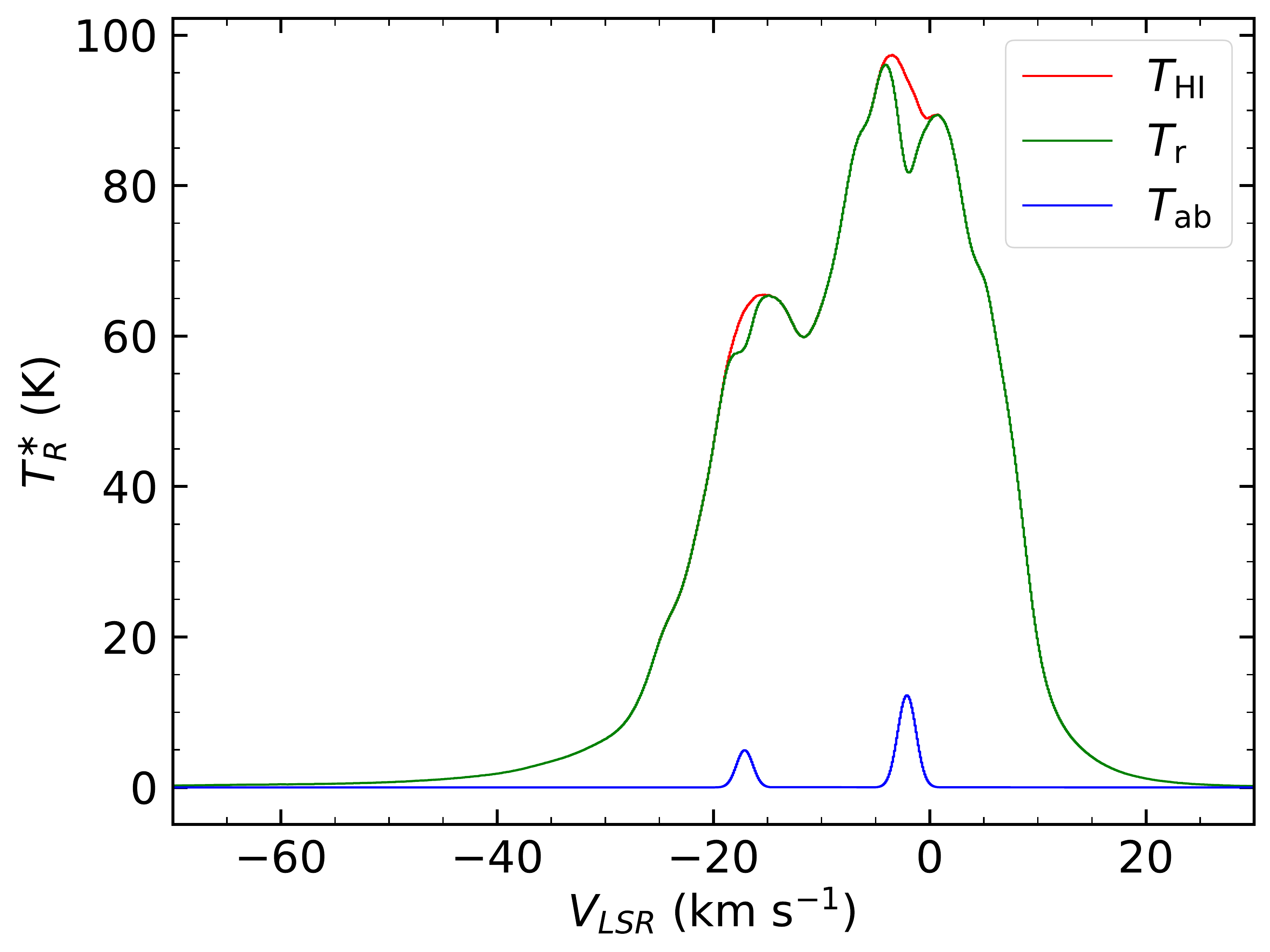}}	
	\caption{HINSA features of Beams 1--7, where the green lines (marked as $T_{\mathrm{r}}$) show the observed spectrum, the red lines (marked as $T_{\mathrm{HI}}$) the recovered spectrum of H \textsc{i}, and the blue lines (marked as $T_{\mathrm{ab}}$) the fitted spectrum of the HINSA features.}
	\label{fig-HINSA-spectrum}
\end{figure*}

We extracted fourteen HINSA features in total from the seven beams (see Figure \ref{fig-HINSA-spectrum}). Their velocity, $V_{\mathrm{LSR}}$, peak intensity, $T_{\mathrm{ab}}$, optical depth, $\tau_0$, and velocity dispersion, $\sigma_{\mathrm{HI}}$, with 1$\sigma$ errors are listed in Table \ref{tab:HINSA-features}. 

The fitted values of $\tau_0$ and $\sigma_{\mathrm{HI}}$ were used to calculate the column density of the HINSA features, $N_{\mathrm{HI}}$ \citep[see][]{Li-Goldsmith2003} as
\begin{equation}\label{equ:NHI}
	N_{\mathrm{HI}} = 1.95 \times 10^{18} \tau_0 T_{\mathrm{ex}} \Delta V,
\end{equation}
where $\Delta V = \sqrt{8\ln(2)}\sigma_{\mathrm{HI}}$. The results of $N_{\mathrm{HI}}$ and the corresponding $T_{\mathrm{ex}}$ with 1$\sigma$ errors are listed in Table \ref{tab:HINSA-features}. The values of $N_{\mathrm{HI}}$ range from $\sim$ 5.3 to 33.1 $\times$ 10$^{18}$ cm$^{-2}$ in the SCER, and from $\sim$ 0.5 to 3.4 $\times$ 10$^{18}$ cm$^{-2}$ in the WCER.

\begin{deluxetable*}{lccccccc}[hbt!]
	\tablecaption{Physical Parameters of the HINSA Features\label{tab:HINSA-features}}	
	\tablehead{
		\colhead{Beams} & \colhead{Index} & \colhead{$V_{\mathrm{LSR}}$} & \colhead{$\sigma_{\mathrm{HI}}$} &  \colhead{$\tau_{\mathrm{HI}}$} & \colhead{$T_{\mathrm{ex}}$} & \colhead{$T_{\mathrm{ab}}$} & \colhead{$N_{\mathrm{HI}}$}  \\
		\colhead{} & \colhead{} & \colhead{(km s$^{-1}$)} & \colhead{(km s$^{-1}$)} & \colhead{} & \colhead{(K)} & \colhead{(K)} & \colhead{(10$^{18}$ cm$^{-2}$)} \\
	\colhead{(1)} & \colhead{(2)} & \colhead{(3)} & \colhead{(4)} & \colhead{(5)} & \colhead{(6)} & \colhead{(7)} & \colhead{(8)}}
	\startdata
		Beam 1	& 1 & $-$18.42 $\pm$ 0.02	&	1.14 $\pm$ 0.02 	&	0.31 $\pm$ 0.01 & 20.30 $\pm$ 0.08	&	18.97 $\pm$ 0.008 & 33.1 $\pm$ 1.1 \\
		& 2 & $-$3.47 $\pm$ 0.02 & 0.85 $\pm$ 0.02 & 0.12 $\pm$ 0.01 & 7.18 $\pm$ 0.18 & 11.89 $\pm$ 0.008 & 3.4 $\pm$ 0.3 \\
		Beam 2 & 1 & $-$17.65 $\pm$ 0.02 & 0.78 $\pm$ 0.02 & 0.07 $\pm$ 0.01 & 20.30 $\pm$ 0.08 & 4.33 $\pm$ 0.006 & 5.3 $\pm$ 0.7 \\
		& 2 & $-$3.83 $\pm$ 0.02 & 0.53 $\pm$ 0.02 & 0.03 $\pm$ 0.01 & 7.18 $\pm$ 0.18 & 2.57 $\pm$ 0.006 & 0.5 $\pm$ 0.2 \\
		Beam 3 & 1 & $-$16.65 $\pm$ 0.02 & 1.46 $\pm$ 0.02 & 0.16 $\pm$ 0.01 & 20.30 $\pm$ 0.08 & 11.25 $\pm$ 0.007 & 21.5 $\pm$ 1.4 \\
		& 2 & $-$2.35 $\pm$ 0.09 & 0.80 $\pm$ 0.09 & 0.04 $\pm$ 0.01 & 5.08 $\pm$ 0.31 & 4.36 $\pm$ 0.007 & 0.8 $\pm$ 0.2 \\
		Beam 4 & 1 & $-$17.63 $\pm$ 0.02 & 0.96 $\pm$ 0.02 & 0.28 $\pm$ 0.01 & 20.30 $\pm$ 0.08 & 17.47 $\pm$ 0.011 & 25.4 $\pm$ 0.9 \\
		& 2 & $-$2.57 $\pm$ 0.09 & 0.78 $\pm$ 0.09 & 0.11 $\pm$ 0.01 & 5.08 $\pm$ 0.31 & 10.88 $\pm$ 0.011 & 2.0 $\pm$ 0.2 \\
		Beam 5 & 1 & $-$18.19 $\pm$ 0.02 & 1.15 $\pm$ 0.02 & 0.21 $\pm$ 0.01 & 20.30 $\pm$ 0.08 & 13.91 $\pm$ 0.006 & 22.8 $\pm$ 1.1 \\
		& 2 & $-$3.23 $\pm$ 0.09 & 0.77 $\pm$ 0.09 & 0.13 $\pm$ 0.01 & 5.08 $\pm$ 0.31 & 13.01 $\pm$ 0.006 & 2.3 $\pm$ 0.3 \\
		Beam 6 & 1 & $-$18.38 $\pm$ 0.02 & 1.24 $\pm$ 0.02 & 0.18 $\pm$ 0.01 & 20.30 $\pm$ 0.08 & 10.06 $\pm$ 0.006 & 20.5 $\pm$ 1.2 \\
		& 2 & $-$2.41 $\pm$ 0.02 & 0.80 $\pm$ 0.02 & 0.10 $\pm$ 0.01 & 4.76 $\pm$ 0.16 & 9.61 $\pm$ 0.006 & 1.7 $\pm$ 0.2 \\
		Beam 7 & 1 & $-$17.19 $\pm$ 0.02 & 0.76 $\pm$ 0.02 & 0.08 $\pm$ 0.01 & 20.30 $\pm$ 0.08 & 4.94 $\pm$ 0.008 & 5.7 $\pm$ 0.7 \\
		& 2 & $-$2.14 $\pm$ 0.02 & 0.83 $\pm$ 0.02 & 0.14 $\pm$ 0.01 & 4.76 $\pm$ 0.16 & 12.23 $\pm$ 0.008 & 2.5 $\pm$ 0.2 
	\enddata
	\tablecomments{Columns 3-–8 present the central velocity, velocity dispersion, optical depth, excitation temperature, peak intensity, and column density of HINSA features.}
\end{deluxetable*}

\subsection{Physical Properties of the Molecular Line Emission}\label{sec-HINSA-compare-with-molecule}

To make a comparison between the HINSA features and the molecular line emission, we resampled the molecular data observed by \citet{Liu+2021} to pixel sizes of $3' \times 3'$, and aligned pixel center of molecular emission with the corresponding FAST beam center.

\begin{figure*}[!htb]
	\centering
	\subfigure[Beam 1]{\includegraphics[height=0.22\textwidth]{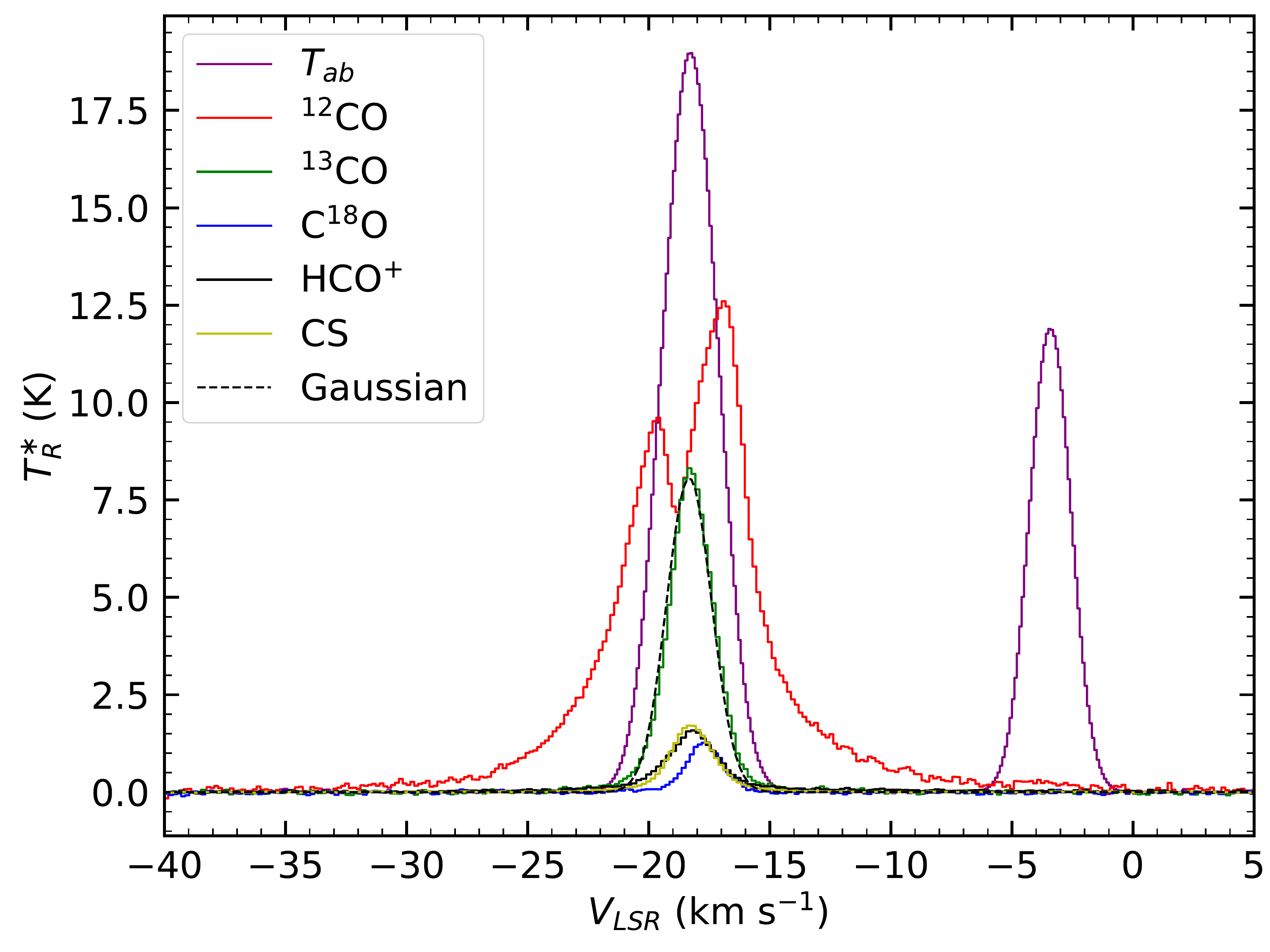}}
	\subfigure[Beam 2]{\includegraphics[height=0.22\textwidth]{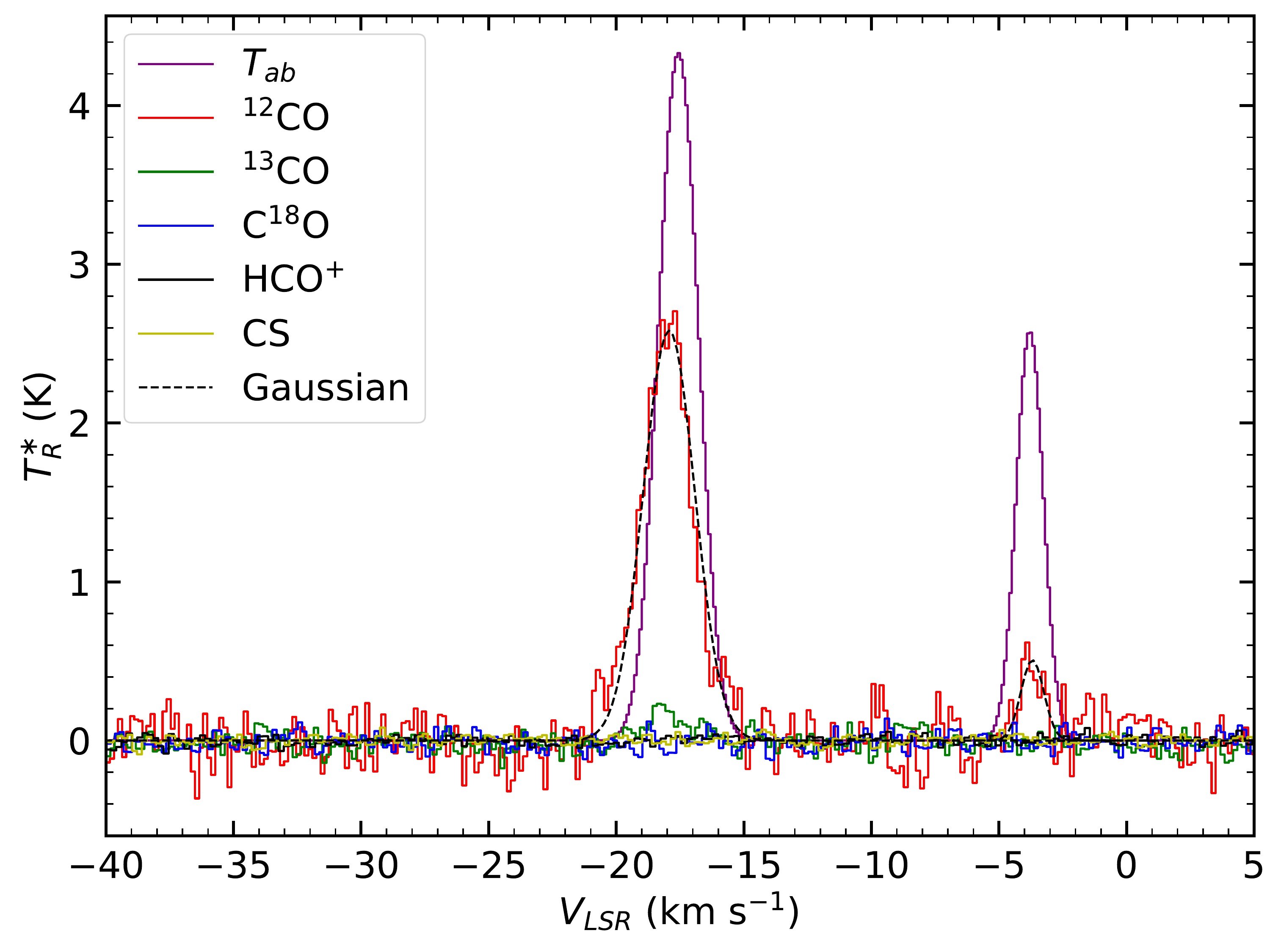}}
	\subfigure[Beam 3]{\includegraphics[height=0.22\textwidth]{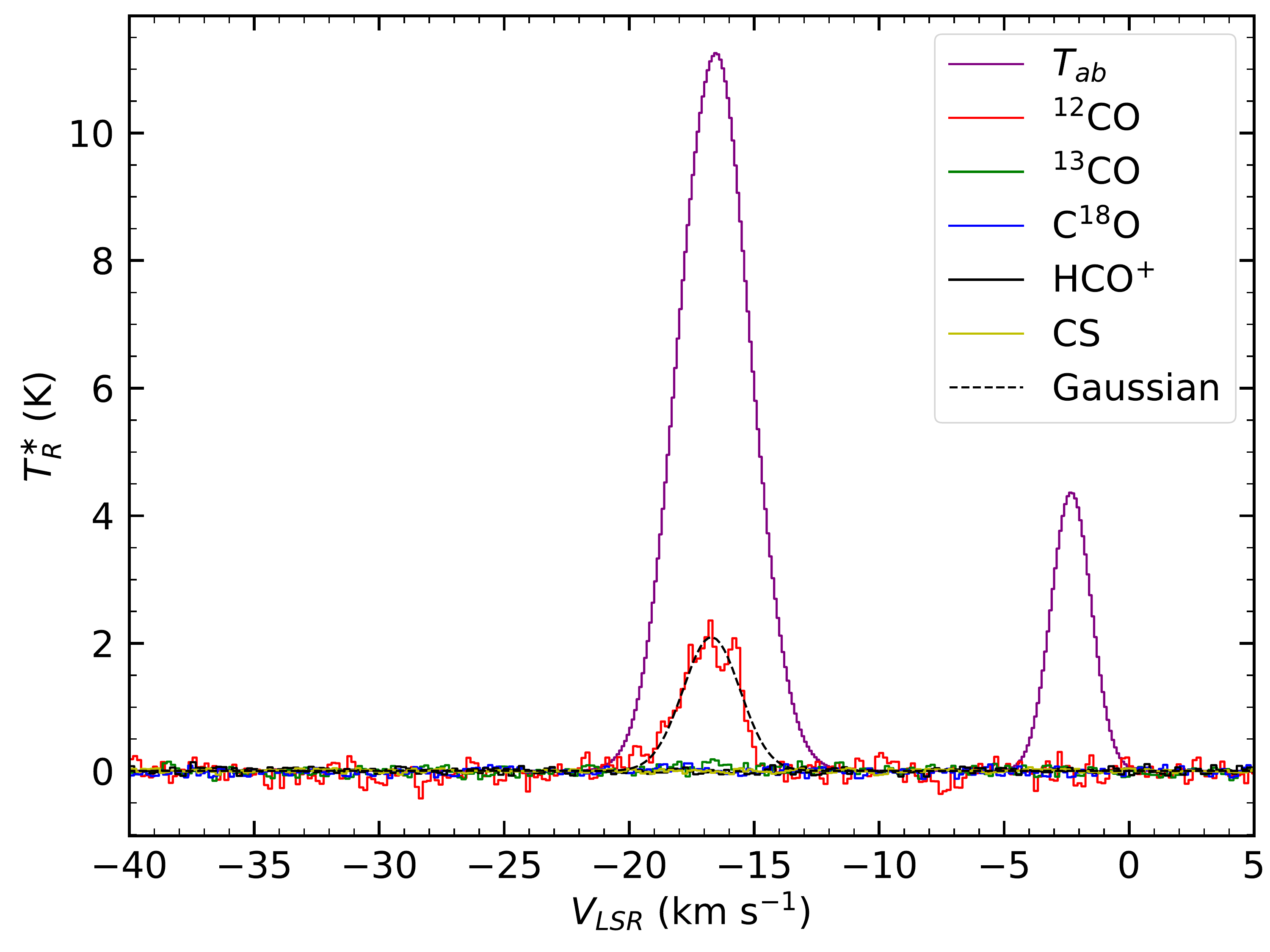}}
	\subfigure[Beam 4]{\includegraphics[height=0.22\textwidth]{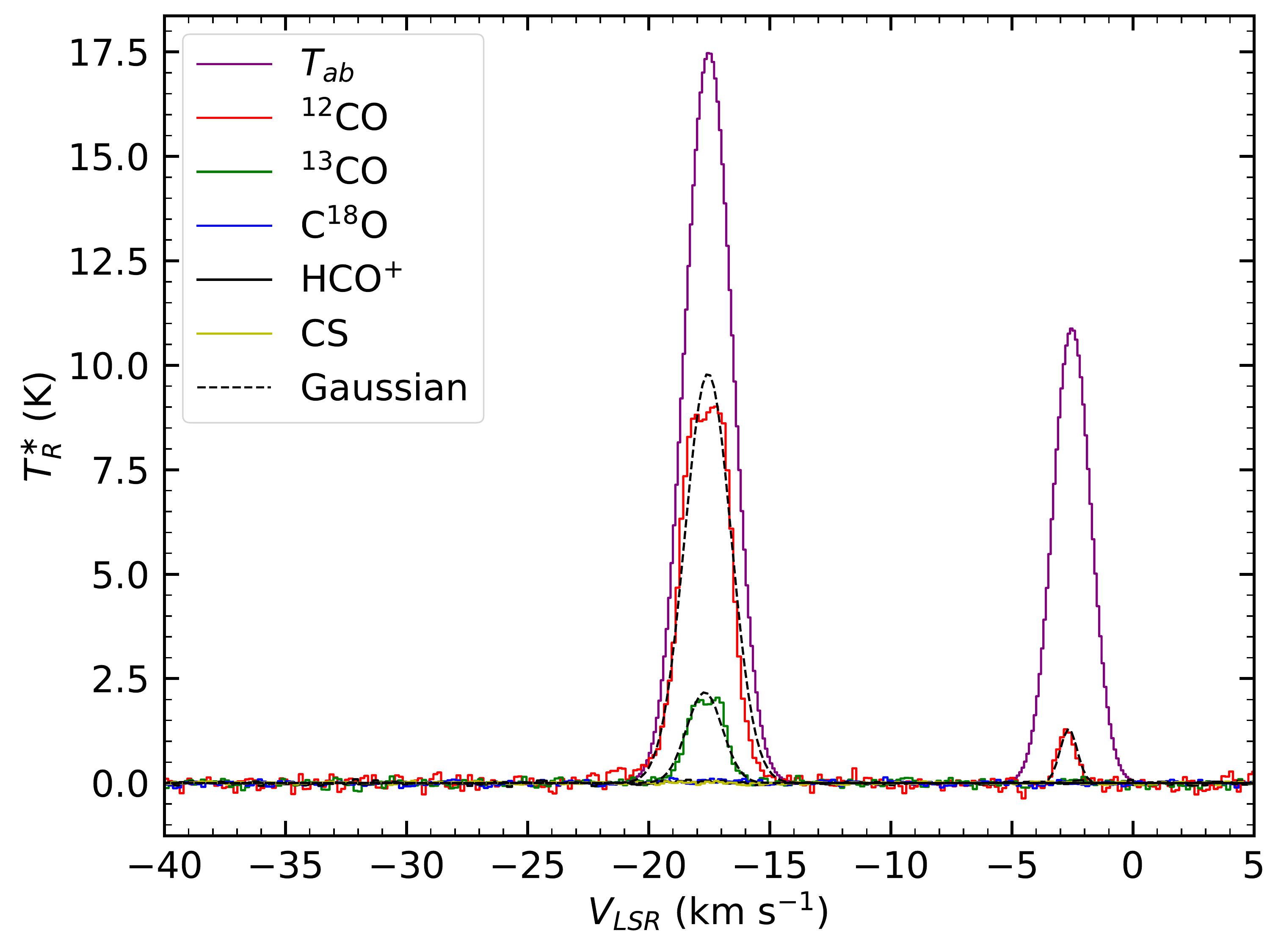}}
	\subfigure[Beam 5]{\includegraphics[height=0.22\textwidth]{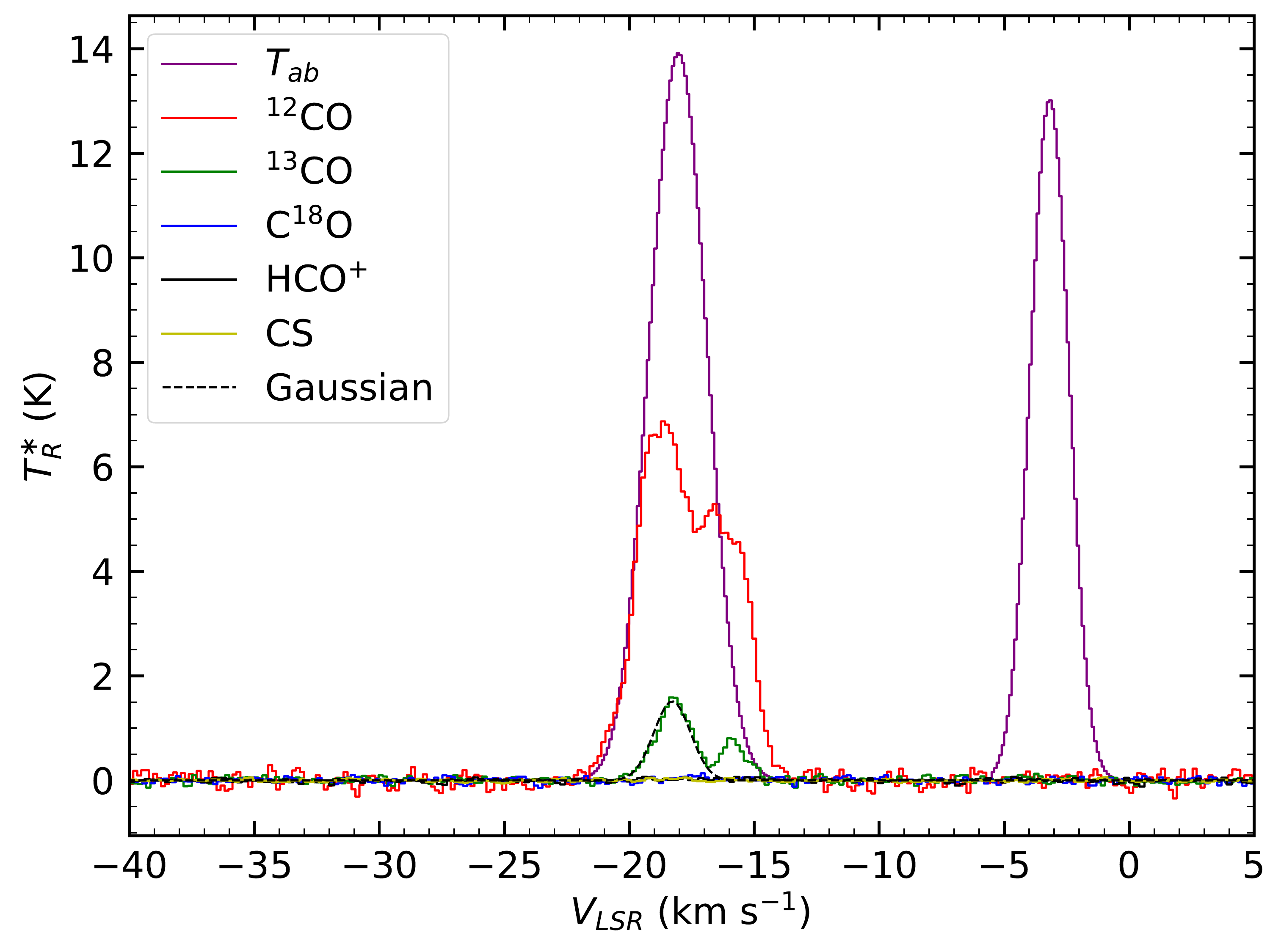}}
	\subfigure[Beam 6]{\includegraphics[height=0.22\textwidth]{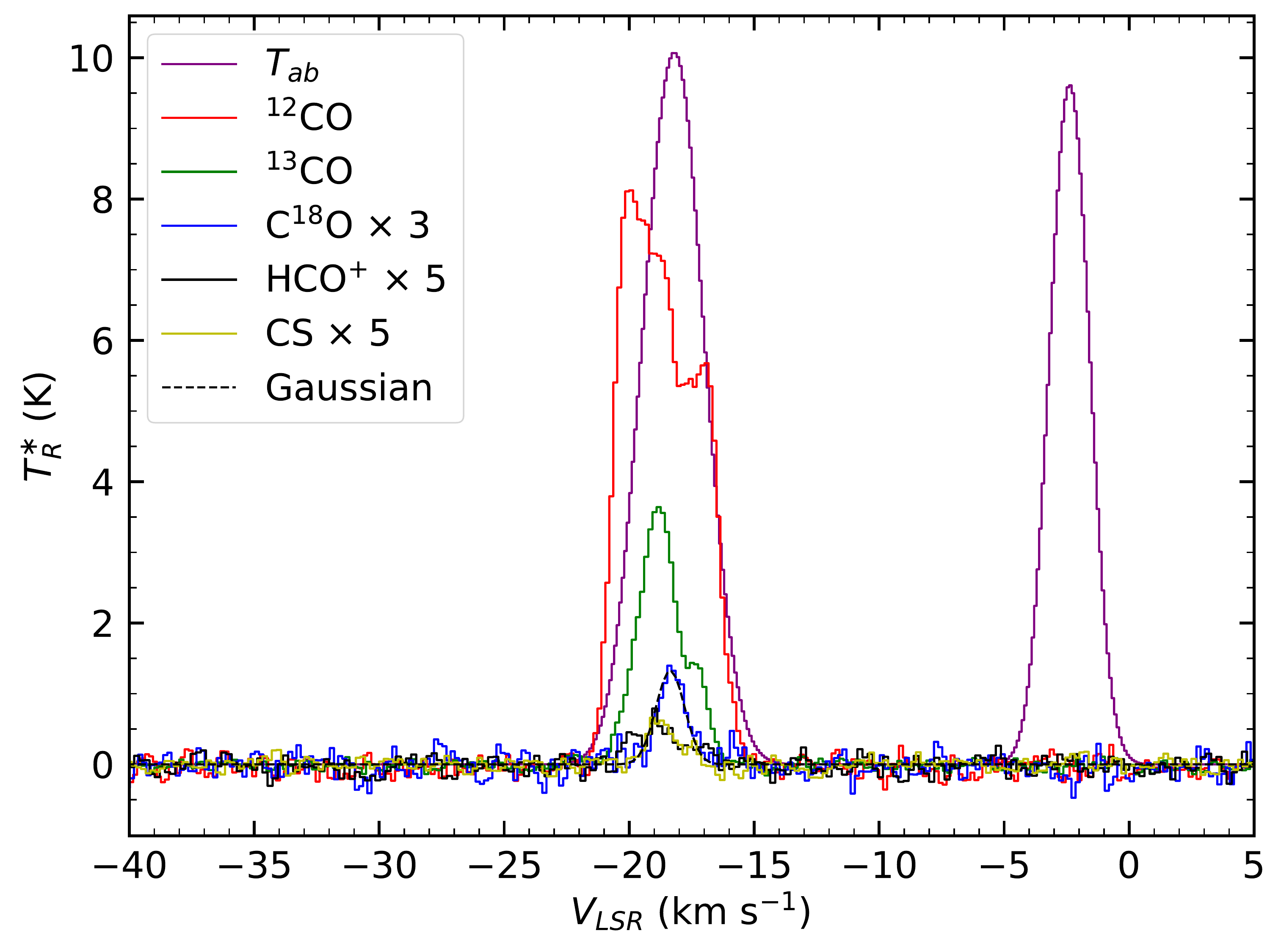}}
	\subfigure[Beam 7]{\includegraphics[height=0.22\textwidth]{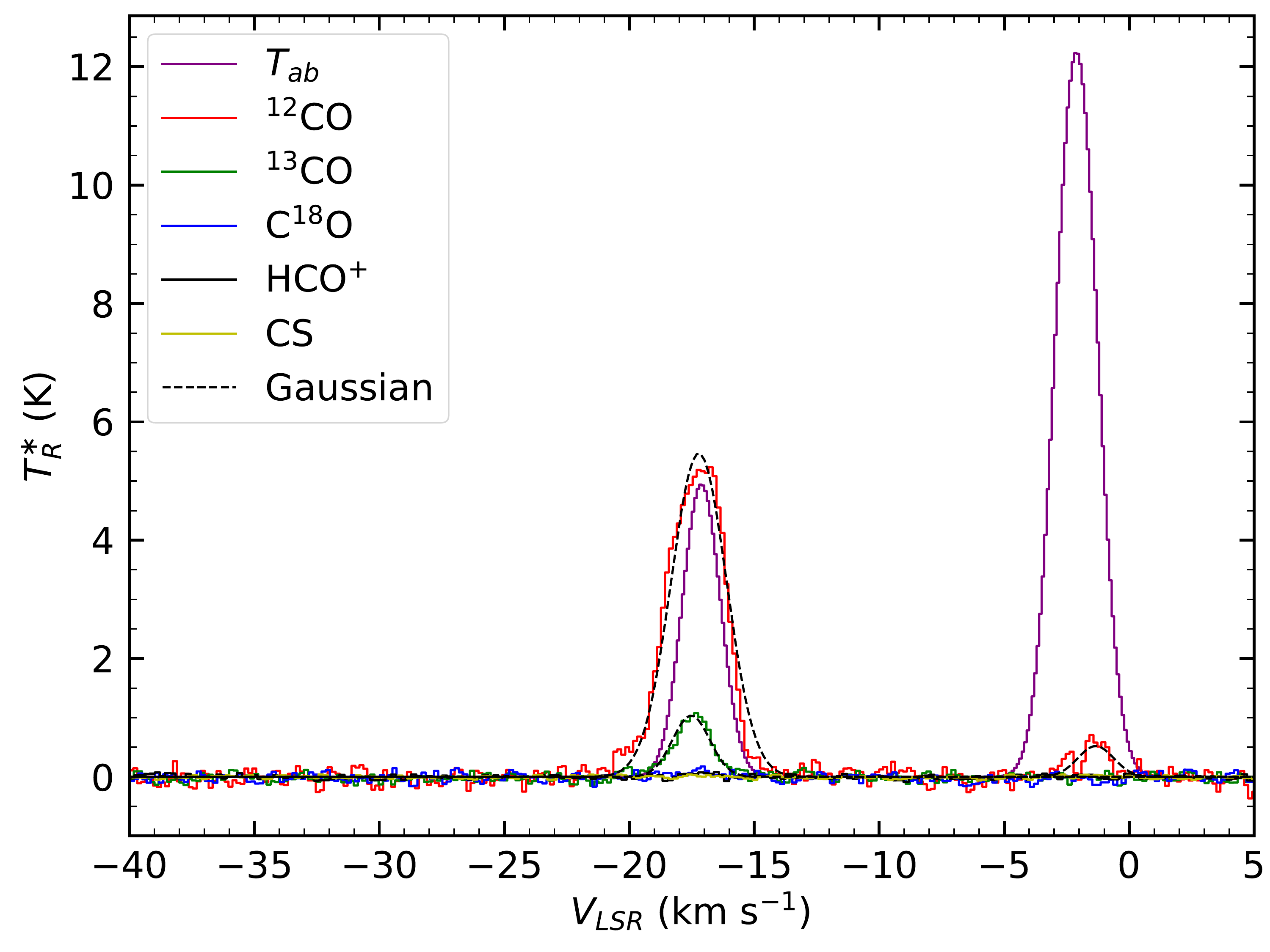}}	
	\caption{HINSA features in Beams 1--7 and the molecular line spectra of the corresponding regions. The red, green, blue, black, and yellow lines show the spectra of $^{12}$CO, $^{13}$CO, C$^{18}$O, HCO$^+$, and CS, respectively, and the purple lines represent the HINSA features. We fitted the profiles of both $^{12}$CO and $^{13}$CO with a Gaussian profile for each component, except for spectra that showed strong self-absorption or were contaminated by other components. For Beam 6, only the line emission of C$^{18}$O was fitted because there were strong self-absorption in the spectra of $^{12}$CO and $^{13}$CO.}
	\label{fig-spectrum-HINSA-CO}
\end{figure*}

Figure \ref{fig-spectrum-HINSA-CO} presents the molecular line emission after resampling, corresponding to each beam as well as the corresponding HINSA features, i.e., $T_{\mathrm{ab}}$. $^{12}$CO emission with a velocity of $\sim$ $-$18 km s$^{-1}$ (i.e., the SCER) is detected in all seven beams, indicating that the detected $^{12}$CO emission in the SCER covers almost the entire observed region of the molecular lines. The line centers and velocity dispersions of the molecular lines are from the Gaussian fits to the $^{12}$CO, $^{13}$CO, and/or C$^{18}$O spectra. We also calculated the column density of all observed molecular species, i.e., $^{12}$CO, $^{13}$CO, C$^{18}$O, HCO$^+$, and CS when these were detected (see Section \ref{column-density-molecular}). The velocity center, velocity dispersion, and column density as well as the corresponding $T_{\mathrm{ex}}$, are listed in Table \ref{tab:molecular-parameters}.

\section{Discussion and and Conclusions}\label{sec-discuss}

\subsection{Association Between the HINSA Features and the Molecular Clouds}\label{sec-HINSA-compare-with-molecule-association}

The association between the HINSA features and the molecular line emissions can be seen in Figure \ref{fig-spectrum-HINSA-CO}. In the SCER (shown as the first entry for each beam in Table \ref{tab:HINSA-features}), the HINSA features are probably associated with the molecular line emission. In the WCER (shown as the second entry in Table \ref{tab:HINSA-features}), no molecular line emission associated with HINSA features is detected toward Beams 1, 3, 4, and 6. The only detected $^{12}$CO emission, corresponding to Beams 2, 5, and 7, is also weak. The HINSA features in the WCER are probably associated with CO-dark or CO-faint gas \citep[e.g.,][]{Wolfire+2010, Tang+2016, Liu+2022}. This is an issue worthy of further observations.

In addition, a further quantitative comparison was conducted regarding the relationship of the center velocity and the non-thermal velocity dispersion between the HINSA features and the molecular gases. 
In the SCER, the differences between the center velocity of the HINSA features (see Table \ref{tab:HINSA-features}) and those of the $^{12}$CO, $^{13}$CO, and/or C$^{18}$O lines are within the velocity resolution of the CO spectra (i.e., $\sim$ 0.2 km s$^{-1}$). This indicates that the HINSA features are associated with the molecular clouds traced by the CO. The non-thermal velocity dispersion of the HINSA features, $\sigma_{\mathrm{nt,HI}} = \sqrt{\sigma_{\mathrm{HI}}^2 - k T_{\mathrm{ex}}/m_{\mathrm{H}}}$, is smaller than that of the $^{12}$CO lines, $\sigma_{\mathrm{nt,^{12}CO}} = \sqrt{\sigma_{\mathrm{^{12}CO}}^2 - k T_{\mathrm{ex}}/m_{\mathrm{^{12}CO}}}$, and is similar to that of the $^{13}$CO lines, $\sigma_{\mathrm{nt,^{13}CO}} = \sqrt{\sigma_{\mathrm{^{13}CO}}^2 - k T_{\mathrm{ex}}/m_{\mathrm{^{13}CO}}}$, except for that in Beam 3, where the value of $\sigma_{\mathrm{nt,HI}}$ is larger than that of $\sigma_{\mathrm{nt,^{12}CO}}$ by $\sim$ 20\% (see $\sigma_{\mathrm{HI}}$ in Table \ref{tab:HINSA-features}, and $\sigma_{\mathrm{^{12}CO}}$ and $\sigma_{\mathrm{^{13}CO}}$ in Table \ref{tab:molecular-parameters}). This result is in agreement with that of \citet{Li-Goldsmith2003}, indicating that the HINSA in the SCER is probably mixed with the gas in cold and well-shielded regions of the molecular clouds. 

In the WCER, the association between the HINSA features and the CO cloud appears to be poor. First, the difference between the center velocity of the HINSA features (see Table \ref{tab:HINSA-features}) and that of $^{12}$CO is larger than in the case of the SCER, and the maximum difference reaches $\sim$ 0.8 km s$^{-1}$ (i.e., in Beam 7). Second, the value of $\sigma_{\mathrm{nt,HI}}$ is larger than that of $\sigma_{\mathrm{nt,^{12}CO}}$. No $^{13}$CO emission is detected toward the regions corresponding to all the seven beams after resampling. In fact, we also do not detect any $^{13}$CO emission in Beam 3 in the SCER, and the value of $\sigma_{\mathrm{nt,HI}}$ in this region is larger than that of $\sigma_{\mathrm{nt,^{12}CO}}$. Therefore, it is probably more accurate to classify Beam 3 in the SCER as a WCER component. The discussion below is based on this reclassification. We also find that the value of $\sigma_{\mathrm{HI}}$ in the WCER (including Beam 3 in the SCER) is comparable to some of the values found by \citet{Tang+2016} for CO-dark clouds. These facts indicate that the HINSA toward the WCER is probably not mixed with CO clouds, but probably associated with more diffuse CO-dark or CO-faint clouds. Further observations of CI, CII, and/or OH may be helpful in clarifying the relationship between the HINSA features and the CO-dark or CO-faint gas \citep[e.g.,][]{Li-Goldsmith2003, Wolfire+2010, Tang+2016} toward the WCER.

\subsection{Abundance of H~\textsc{i} the HINSA Features}\label{sec-HINSA-compare-with-molecule-relation}

The fractional abundance of HINSA, $X_{\mathrm{HI}}$, based on the column density is valuable. In this case $X_{\mathrm{HI}} = N_{\mathrm{HI}}/N_{\mathrm{H}}$, where $N_{\mathrm{HI}}$ is the column density of the HINSA, and $N_{\mathrm{H}} = N_{\mathrm{HI}} + 2N_{\mathrm{H_2}}$. $N_{\mathrm{H_2}}$ is the column density of H$_2$ in the molecular cloud, which can be traced by $^{12}$CO, $^{13}$CO, C$^{18}$O, HCO$^+$, and CS, i.e., $N^{\mathrm{^{12}CO}}_{\mathrm{H_2}}$, $N^{\mathrm{^{13}CO}}_{\mathrm{H_2}}$, $N^{\mathrm{C^{18}O}}_{\mathrm{H_2}}$, $N^{\mathrm{HCO^+}}_{\mathrm{H_2}}$, and $N^{\mathrm{CS}}_{\mathrm{H_2}}$ (see Table \ref{tab:molecular-parameters}).

We compared the values of $N^{\mathrm{^{12}CO}}_{\mathrm{H_2}}$ and $N^{\mathrm{^{13}CO}}_{\mathrm{H_2}}$ within all regions where we simultaneously detected $^{12}$CO and $^{13}$CO line emission (i.e., Beams 1, 2, and 4--7 of the SCER). The results show that the value of $N^{\mathrm{^{13}CO}}_{\mathrm{H_2}}$ is larger than that of $N^{\mathrm{^{12}CO}}_{\mathrm{H_2}}$ by a factor of $\lesssim$ 3, indicating that $N^{\mathrm{^{13}CO}}_{\mathrm{H_2}}$ is a better indicator of the column density of the molecular clouds. The values of $N^{\mathrm{^{12}CO}}_{\mathrm{H_2}}$ are probably underestimated due to optical depth effects on the $^{12}$CO spectra. In addition, the species tracing higher densities, i.e., C$^{18}$O, HCO$^+$, and CS, are detected only in Beams 1 and 6. In Beam 1, the ratios of the column density traced by $^{13}$CO, $N^{\mathrm{^{13}CO}}_{\mathrm{H_2}}$, to those traced by the three dense gas tracers, $N^{\mathrm{C^{18}O}}_{\mathrm{H_2}}$, $N^{\mathrm{HCO^+}}_{\mathrm{H_2}}$, and $N^{\mathrm{CS}}_{\mathrm{H_2}}$, are $\sim$ 0.92, $\sim$ 1.48, and $\sim$ 0.78, respectively, indicating that $N^{\mathrm{^{13}CO}}_{\mathrm{H_2}}$ is a good measure of the column density of the molecular cloud. In Beam 6, the dilution of the dense gas tracers is relatively stronger (see Figure \ref{fig-12co-moment}(b)). Nevertheless, $N^{\mathrm{^{13}CO}}_{\mathrm{H_2}}/N^{\mathrm{C^{18}O}}_{\mathrm{H_2}} \sim 1.2$ also indicates that $^{13}$CO is a good tracer of the molecular cloud. Therefore, we used $N^{\mathrm{^{13}CO}}_{\mathrm{H_2}}$ to represent the column density of molecular clouds in the SCER.

\begin{figure*}[!htb]
	\centering
	\subfigure[$N^{\mathrm{^{12}CO}}_{\mathrm{H_2}}$ .VS. $N^{\mathrm{^{13}CO}}_{\mathrm{H_2}}$]{\includegraphics[height=0.3\textwidth]{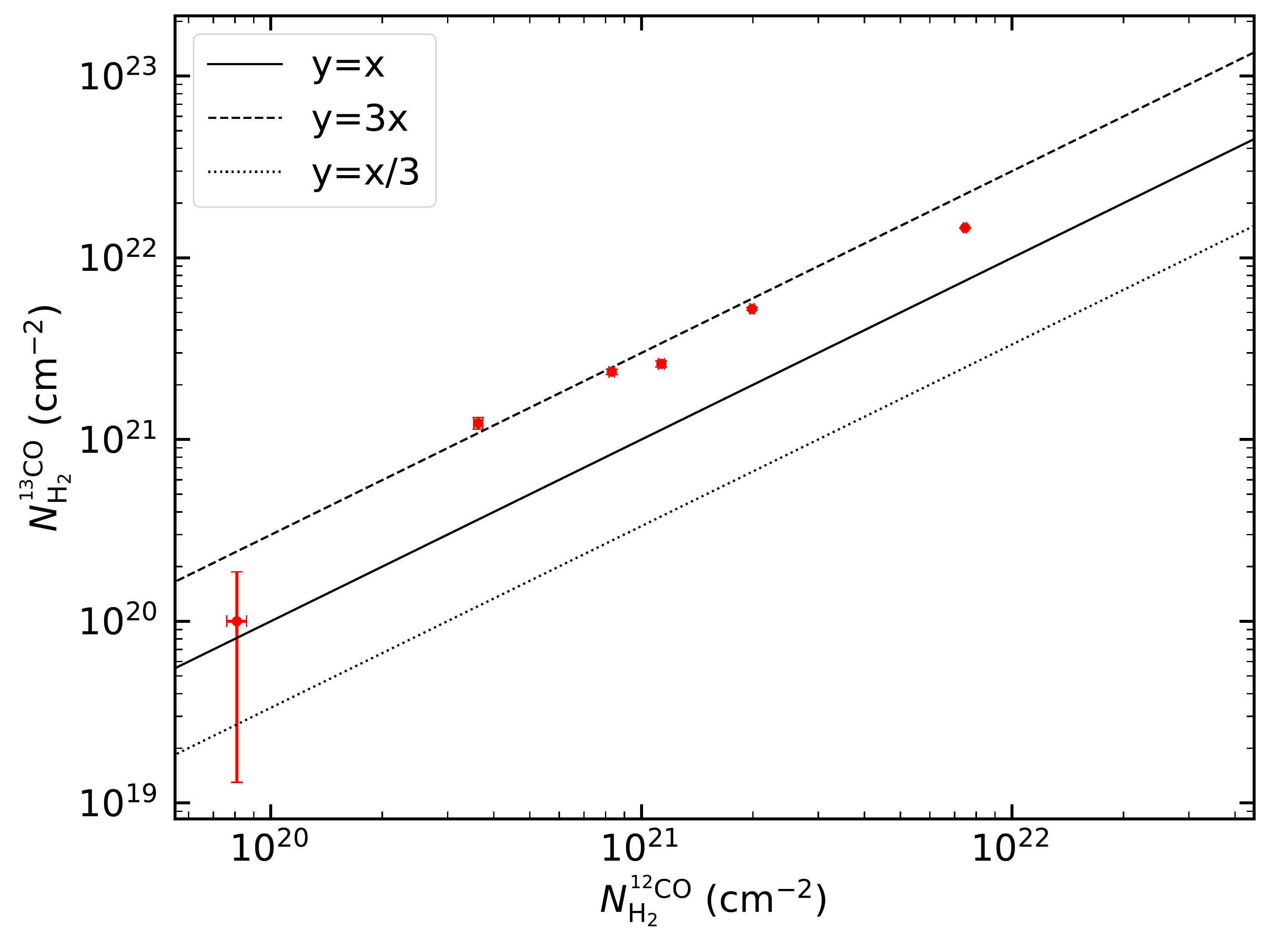}}
	\subfigure[$N^{\mathrm{^{13}CO}}_{\mathrm{H_2}}$ (or $N^{\mathrm{^{12}CO}}_{\mathrm{H_2}}$) .VS. $N_{\mathrm{HI}}$]{\includegraphics[height=0.3\textwidth]{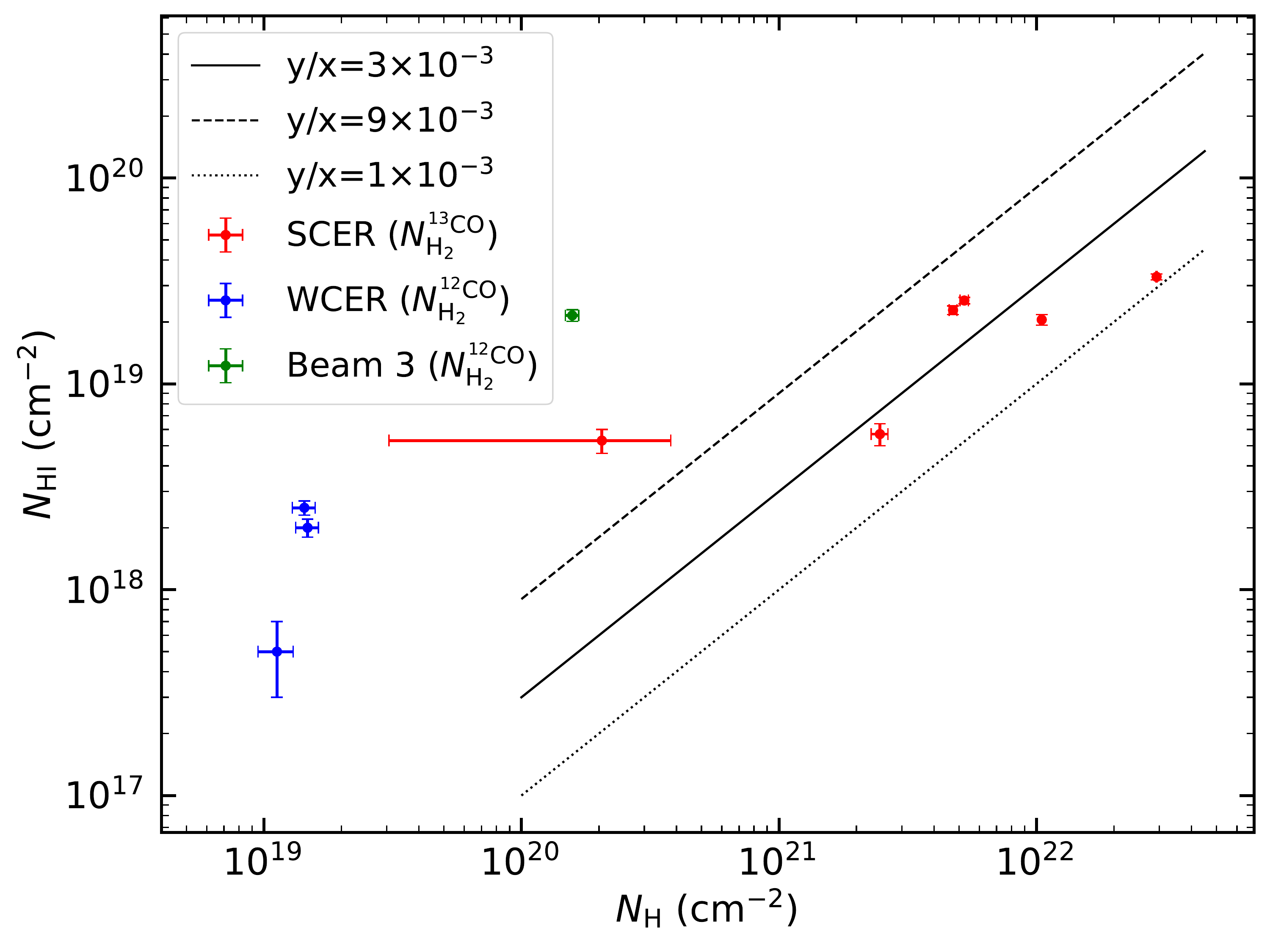}}
	\caption{Left: comparison of $N^{\mathrm{^{12}CO}}_{\mathrm{H_2}}$ and $N^{\mathrm{^{13}CO}}_{\mathrm{H_2}}$ in the SCER, where $^{13}$CO is not detected in Beam 3. Right: comparison of the column density of HINSA, $N_{\mathrm{HI}}$, and of the total gas, $N_{\mathrm{H}}$. The red points represent the results corresponding to the SCER (except for Beam 3), where the column density of the molecular clouds is traced by $^{13}$CO. The blue points present the results corresponding to the WCER, where the column density of the molecular clouds is traced by $^{12}$CO. The green points indicate the result corresponding to Beam 3 in the SCER (which may be more suitably classified as a WCER component), where the column density of the molecular clouds is traced by $^{12}$CO.}
	\label{fig-columndensity}
\end{figure*}

The abundance of HINSA, $X_{\mathrm{HI}}$, in most cases of the SCER (i.e., Beams 1 and 4--7), ranges from $\sim$ $1.1 \times 10^{-3}$ to $\sim$ $4.8 \times 10^{-3}$ (see the red points in Figure \ref{fig-columndensity}(b)). The one exception is the value of $X_{\mathrm{HI}}$ is $\sim$ $2.6 \times 10^{-2}$ in Beam 2, where the value of $N^{\mathrm{^{13}CO}}_{\mathrm{H_2}}$ is one order of magnitude smaller than those in the other five beams (i.e., Beams 1 and 4--7), and the relative error of $N^{\mathrm{^{13}CO}}_{\mathrm{H_2}}$ is large (see Figure \ref{fig-columndensity}(b) and Table \ref{tab:molecular-parameters}). The abundances of HINSA in Beams 1 and 4--7 are similar to those toward optically selected dark clouds and molecular cores \citep[i.e., with an average value of $\sim$ 1.5 $\times$ 10$^{-3}$,][]{Li-Goldsmith2003, Krco-Goldsmith2010}. The HINSA abundances in the SCER (including Beam 2) fall into the HINSA abundance range toward PGCC sources (i.e., from 5.1 $\times$ 10$^{-4}$ to 1.3 $\times$ 10$^{-2}$) of \cite{Tang+2020}, but larger than the HINSA abundances derived from other PGCC sources (i.e., $\sim$ 3 $\times$ 10$^{-4}$ varied by a factor of $\sim$ 3) by \cite{Liu+2022}.

The HINSA abundance toward Beam 1 is consistent with the abundance of H~\textsc{i} in the H~\textsc{i} wind (i.e., the ratio of the column density of the H~\textsc{i} wind to that of the total outflowing gas) in the same beam. This fact indicates that HINSA is probably a physical bridge that connects H~\textsc{i} winds and molecular outflows.

We also find that the value of $X_{\mathrm{HI}}$ in Beam 1 is smaller than the values obtained from the surrounding beams (i.e., Beams 2--7; see Figure \ref{fig-12co-moment}(a)). This structure seems to be a ring, but unlike that observed by \citet{Zuo+2018}, the abundance of HINSA, $X_{\mathrm{HI}}$, diminishes toward the center of the cloud. One of the reason for this could be that the central high-mass zero-age main sequence B3 star \citep[see][]{Torrelles+1992} ionizes the H~\textsc{i} detected in Beam 1 in the SCER, which is located in an H~\textsc{ii} region \citep[see][]{Dewangan2019}. This then decreases the abundance of H~\textsc{i}. The ionization can be studied using radio recombination lines \citep[e.g.,][]{Zhang+2014, Zhang+2021}. 

The mean densities of H~\textsc{i}, $\bar{n}_{\mathrm{HI}}$, and total proton, $\bar{n}_{\mathrm{H}}$, in Beam 1 are $\sim$ 7.2 and $\sim$ 6.1 $\times$ 10$^{3}$ cm$^{-3}$, respectively, assuming that the cloud diameter is $\sim$ 1.5 pc, corresponding to the diameter being the HPBW of FAST. We assume that the density profile has a constant density core surrounded by an envelope of the form
\begin{equation}\label{equ:density-profile}
n(r) = \frac{2n_0}{1+(r/r_0)^2},	
\end{equation}
where $n_0$ and $r_0$ (set to 10$^{17}$ cm) are the central density and the radius of the constant density core. In this case, $n_0 \sim 9.4\, \bar{n}_{\mathrm{H}}$ would be larger than that of model 1 in \citet{Goldsmith+2007}, indicating that the density is high enough to initiate the conversion of H~\textsc{i} $\rightarrow$ H$_2$.
Following the model developed by \citet{Goldsmith+2007}, the time that has elapsed since the material was UV irradiated is $\sim$ 2 $\times$ 10$^{5}$ yr (see Section \ref{sec:timescale}), during which H~\textsc{i} could convert to H$_2$. This time is consistent with the timescale of the star formation in this region (i.e., $\sim$ 2 $\times$ 10$^{5}$ yr) derived from near-infrared H$_2$ knots by \citet{Chen+2003}. The model of \citet{Goldsmith+2007} also implies another possible reason why $X_{\mathrm{HI}}$ diminishes toward the center of the cloud, i.e., the conversion to molecular form occurs more rapidly in the higher-density inner region.

In the WCER, a comparison of $N_{\mathrm{HI}}$ and $N_{\mathrm{H}}$ (only for the regions where CO emission is detected) is plotted as blue and green points in Figure \ref{fig-columndensity}(b). The abundance of the HINSA features, $X_{\mathrm{HI}}$, is $\sim$ 0.04, $\sim$ 0.14, and $\sim$ 0.17 in Beams 2, 4, and 7, respectively. The value of $X_{\mathrm{HI}}$ is $\sim$ 0.14 in Beam 3 in the SCER where $^{13}$CO emission is not detected. The high abundances are most likely caused by an underestimation of the column density of the molecular clouds, because the associated molecular clouds (including the region in Beam 3 in the SCER) are probably CO-dark or CO-faint gas \citep[e.g.,][]{Bolatto+2013, Tang+2016}. 

The comparison of the HINSA features and the molecular line emission indicates that the HINSA is probably mixed with the gas in cold and well-shielded CO clouds in which $^{13}$CO emission is detected. There are HINSA features that have no detected $^{13}$CO and/or $^{12}$CO emission counterparts, meaning they are probably associated with CO-dark or CO-faint gas. In addition, HINSA is a promising physical bridge to connect the H~\textsc{i} winds with molecular outflows. There is a possible ring where the abundance of HINSA diminishes toward the center of the cloud.

\appendix

\section{Extracting HINSA}\label{sec-HINSA-extract}

The method used here to extract HINSA is based on the methodology developed by \citet{Liu+2022}. The first step of this method is to obtain the initial values of the optical depth, $\tau_i$, velocity dispersion, $\sigma_i$, and central velocity, $v_i$. $\tau_i$ is initially set to 0.1. Obtaining the initial values of $v_i$ and $\sigma_i$ requires knowledge derived from the corresponding high-sensitivity CO spectrum, including the central velocity, $v_\mathrm{{c,m}}$, velocity dispersion, $\sigma_{\mathrm{m}}$, and excitation temperature, $T_{\mathrm{ex}}$.
To extract HINSA as much as possible, we assume that every component of CO emission (i.e., Peaks 1--4) corresponds to a HINSA feature. The initial value of $v_i$ equals $v_\mathrm{{c,m}}$, and the initial value of $\sigma_i$ reads as
\begin{equation}\label{equ:sigma_i}
	\sigma_i = \sqrt{\sigma_{\mathrm{m}}^2 + \frac{k T_{\mathrm{ex}}}{m_{\mathrm{H}}} - \frac{kT_{\mathrm{ex}}}{m_{\mathrm{CO}}}},
\end{equation}
where $k$ is the Boltzmann constant, $m_{\mathrm{H}}$ the mass of atomic hydrogen, and $m_{\mathrm{CO}}$ the mass of the $^{12}$CO molecule for Peaks 1--3 and of the $^{13}$CO molecule for Peak 4.
 
The optical depth of a HINSA feature can be written \citep{Krco+2008} as
\begin{equation}\label{equ:tau_v}
	\tau(v) = \sum_{m-1}^{i = 0} \displaystyle \tau_i e^{\displaystyle- \frac{(v - v_i)^2}{2\sigma_i^2}},
\end{equation}
where $\tau_i$, $\sigma_i$, and $v_i$ are the initial values of the optical depth, velocity dispersion, and central velocity of the $i$-th component of the HINSA features, respectively. 

The H~\textsc{i} emission free from self-absorption, $T_{\mathrm{HI}}$, and the self-absorption of H~\textsc{i}, $T_{\mathrm{ab}}$, satisfies \citep{Li-Goldsmith2003}
\begin{equation}\label{equ:Tab-THI-Tr}
	T_{\mathrm{ab}} = T_{\mathrm{HI}} - T_{\mathrm{r}} = \left[p T_{\mathrm{HI}} + \left(T_\mathrm{c} - T_{\mathrm{ex}} \right)\left(1 - \tau_\mathrm{f}\right)\right](1 - e^{-\tau}),
\end{equation}
where $p$ is a parameter representing the proportion of the background H~\textsc{i} optical depth, $T_{\mathrm{r}}$ is the observed H~\textsc{i} spectrum, $T_\mathrm{c}$ is the H~\textsc{i} continuum brightness temperature, $T_{\mathrm{ex}}$ is the excitation temperature (derived from the spectrum of $^{12}$CO), and $\tau_\mathrm{f}$ is the foreground H~\textsc{i} optical depth. 

Following the treatment of \citet{Liu+2022}, $p$ was set to unity, and then $\tau_f = 0$. $T_{\mathrm{ex}}$ and $T_c$ were simultaneously ignored. $T_{\mathrm{HI}}$ then can be expressed as
\begin{equation}\label{equ:THI-1}
	T_{\mathrm{HI}} = T_{\mathrm{r}} e^{\tau},
\end{equation}
which can be rewritten as
\begin{equation}\label{equ:THI-2}
	T_{\mathrm{HI}} = T_{\mathrm{r}} + (1 - e^{\tau}) T^f_{\mathrm{HI,smooth}},
\end{equation}
where $T^f_{\mathrm{HI,smooth}}$ is the polynomial fit of $T_{\mathrm{HI}}$. 

$\tau_i$, $\sigma_i$, and $v_i$ \citep{Krco+2008} could be obtained by minimizing $\mathscr{R}$
\begin{equation}\label{equ:R}
	\mathscr{R} \cong \sum^{N}_{j = 1} (T''_{\mathrm{HI}})_j^2 \Delta_{\mathrm{ch}},
\end{equation}
where
\begin{equation}\label{equ:THI-difference}
	T''_{\mathrm{HI}} = \frac{T_{\mathrm{HI}, j + 1} + T_{\mathrm{HI}, j - 1} - 2 T_{\mathrm{HI}, j }}{\Delta_{\mathrm{ch}}^2},
\end{equation}
and where $\Delta_{\mathrm{ch}}$ is the channel width of the H~\textsc{i} spectrum and $j$ the index of the channel.
Similar to \citet{Liu+2022}, the L-BFGS-B algorithm from Scipy \footnote{\url{https://www.scipy.org/}}\citep{2020SciPy-NMeth} was used to obtain $\tau_i$, $\sigma_i$, and $v_i$. For more details of the methodology of extracting HINSA, see \citet{Liu+2022}. The final values of the optical depth, velocity dispersion, and central velocity of the HINSA features are denoted $\tau_0$, $\sigma_{\mathrm{HI}}$, and $V_{\mathrm{LSR}}$, respectively.


\section{Column Densities of the Molecules}\label{column-density-molecular}

For the tracer of $^{13}$CO, under local
thermal equilibrium, the column density of H$_2$, $N^{\mathrm{^{13}CO}}_{\mathrm{H_2}}$, is \citep{Snell+1984}
\begin{equation}\label{equ:N-13CO}
	N^{\mathrm{^{13}CO}}_{\mathrm{H_2}} = 2.3 \times 10^{19} \frac{T_{\mathrm{ex}}}{e^{-5.29/T_{\mathrm{ex}}}} \int \frac{e^{\tau_{\mathrm{^{13}CO}}}}{1 - e^{-\tau_{\mathrm{^{13}CO}}}} T_{\mathrm{mb}}^{^{13}\mathrm{CO}} dv,
\end{equation}
where $X_{\mathrm{^{13}CO}}$ = [$^{13}$CO]/[H$_2$]  = $2 \times 10^{-6}$ is used \citep{Dickman1978}. In the third term, the optical depth ($\tau_{\mathrm{^{13}CO}}$) correction, $e^{\tau_{\mathrm{^{13}CO}}}/(1 - e^{-\tau_{\mathrm{^{13}CO}}})$, is considered only when the values of $T_{\mathrm{mb}}^{^{13}\mathrm{CO}}$ are larger than three times the corresponding rms noise. In this term, the optical depth of $^{13}$CO is
\begin{equation}\label{equ:tau-13co}
\tau_{\mathrm{^{13}CO}} = -\ln\left[1 - \frac{T_{\mathrm{mb}}^{^{13}\mathrm{CO}}/5.29}{1/(e^{5.29}/T_{\mathrm{ex}} - 1) - 0.16}\right].
\end{equation}

$\tau_{\mathrm{^{13}CO}}$ can be used to calculate the optical depth of $^{12}$CO, $\tau_{\mathrm{^{12}CO}}$, by multiplying the abundance ratio,
[$^{12}$CO]/[$^{13}$CO]~$\sim$ 50. The column density of H$_2$ traced by $^{12}$CO is therefore
\begin{equation}\label{equ:N-12co}
	N^{\mathrm{^{12}CO}}_{\mathrm{H_2}} = 4.2 \times 10^{17} \frac{T_{\mathrm{ex}}}{e^{-5.53/T_{\mathrm{ex}}}} \int \frac{e^{\tau_{\mathrm{^{12}CO}}}}{1 - e^{-\tau_{\mathrm{^{12}CO}}}} T_{\mathrm{mb}}^{^{12}\mathrm{CO}} dv,
\end{equation}
where $X_{\mathrm{^{12}CO}}$ = [$^{12}$CO]/[H$_2$] = $10^{-4}$ is used in this work \citep{Snell+1984}. If $^{13}$CO emission is not detected, the optical depth correction in the third term is set to unity.

Similarly, the column density of H$_2$ traced by C$^{18}$O is
\begin{equation}\label{equ:N-c18o}
	N^{\mathrm{C^{18}O}}_{\mathrm{H_2}} = 2.4 \times 10^{20} \frac{T_{\mathrm{ex}}}{e^{-5.27/T_{\mathrm{ex}}}} \int \frac{e^{\tau_{\mathrm{C^{18}O}}}}{1 - e^{-\tau_{\mathrm{C^{18}O}}}} T_{\mathrm{mb}}^{\mathrm{C^{18}O}} dv,
\end{equation}
with
\begin{equation}\label{equ:tau-c18o}
	\tau_{\mathrm{C^{18}O}} = -\ln\left[1 - \frac{T_{\mathrm{mb}}^{\mathrm{C^{18}O}}/5.27}{1/(e^{5.27}/T_{\mathrm{ex}} - 1) - 0.17}\right],
\end{equation}
where $X_{\mathrm{C^{18}O}}$ = [C$^{18}$O]/[H$_2$] = $2 \times 10^{-7}$ is used \citep{Garden+1991}. For HCO$^+$ \citep{Yang+1991} we have
\begin{equation}\label{equ:N-HCO}
	N^{\mathrm{HCO^+}}_{\mathrm{H_2}} = 1.87 \times 10^{19} \frac{T_{\mathrm{ex}}}{1 - e^{-4.28/T_{\mathrm{ex}}}} \int \frac{e^{\tau_{\mathrm{HCO^+}}}}{1 - e^{-\tau_{\mathrm{HCO^+}}}} T_{\mathrm{mb}}^{\mathrm{HCO^+}} dv,
\end{equation}
with
\begin{equation}\label{equ:tau-HCO}
	\tau_{\mathrm{HCO^+}} = -\ln\left[1 - \frac{T_{\mathrm{mb}}^{\mathrm{HCO^+}}/4.28}{1/(e^{4.28}/T_{\mathrm{ex}} - 1) - 0.26}\right],
\end{equation}
where $X_{\mathrm{HCO^+}}$ = [HCO$^+$]/[H$_2$] = $10^{-8}$ is used \citep{Turner+1997}. Finally, for CS we have \citep{Liu+2021}
\begin{equation}\label{equ:N-CS}
	N^{\mathrm{CS}}_{\mathrm{H_2}} = 1.89 \times 10^{20} \frac{T_{\mathrm{ex}}}{e^{-4.70/T_{\mathrm{ex}}}} \int \frac{e^{\tau_{\mathrm{CS}}}}{1 - e^{-\tau_{\mathrm{CS}}}} T_{\mathrm{mb}}^{\mathrm{CS}} dv,
\end{equation}
with
\begin{equation}\label{equ:tau-CS}
	\tau_{\mathrm{CS}} = -\ln\left[1 - \frac{T_{\mathrm{mb}}^{\mathrm{CS}}/4.70}{1/(e^{4.70}/T_{\mathrm{ex}} - 1) - 0.22}\right],
\end{equation}
where $X_{\mathrm{CS}}$ = [CS]/[H$_2$] = 10$^{-9}$ is used \citep{Tatematsu+1998}. $N^{\mathrm{^{12}CO}}_{\mathrm{H_2}}$, $N^{\mathrm{^{13}CO}}_{\mathrm{H_2}}$, $N^{\mathrm{C^{18}O}}_{\mathrm{H_2}}$, $N^{\mathrm{HCO^+}}_{\mathrm{H_2}}$, and $N^{\mathrm{CS}}_{\mathrm{H_2}}$ are listed in Table \ref{tab:molecular-parameters}.

\setcounter{table}{0}
\renewcommand{\thetable}{A\arabic{table}}
\begin{deluxetable*}{lccccccccccc}[hbt!]
	\tabletypesize{\tiny}
	\tablecaption{Physical Parameters Traced by the Molecules	\label{tab:molecular-parameters}}
	\setlength{\tabcolsep}{2.0pt}
	\tablehead{
		\\
		\colhead{Beams} & \colhead{Index} & \colhead{$V_{\mathrm{^{12}CO}}$} & \colhead{$V_{\mathrm{^{13}CO}}$} & \colhead{$\sigma_{\mathrm{^{12}CO}}$} & \colhead{$\sigma_{\mathrm{^{13}CO}}$} & \colhead{$T_{\mathrm{ex}}$} & \colhead{$N^{\mathrm{^{12}CO}}_{\mathrm{H_2}}$} & \colhead{$N^{\mathrm{^{13}CO}}_{\mathrm{H_2}}$} & \colhead{$N^{\mathrm{C^{18}O}}_{\mathrm{H_2}}$} & \colhead{$N^{\mathrm{HCO^+}}_{\mathrm{H_2}}$} &
		\colhead{$N^{\mathrm{CS}}_{\mathrm{H_2}}$} \\
		\colhead{} & \colhead{} & \colhead{(km s$^{-1}$)} & \colhead{(km s$^{-1}$)} & \colhead{(km s$^{-1}$)} & \colhead{(km s$^{-1}$)} & \colhead{(K)} & \colhead{(10$^{20}$ cm$^{-2}$)} & \colhead{(10$^{21}$ cm$^{-2}$)} & \colhead{(10$^{21}$ cm$^{-2}$)} & \colhead{(10$^{21}$ cm$^{-2}$)} & \colhead{(10$^{21}$ cm$^{-2}$)} }
	\startdata
	Beam 1	& 1 & ... &	$-$18.33 $\pm$ 0.04 &... & 0.93 $\pm$ 0.04	& 20.30 $\pm$ 0.08 & 74.72 $\pm$ 0.73 & 14.62 $\pm$ 0.08	&	15.84 $\pm$ 0.53 & 10.29 $\pm$ 0.13 & 21.26 $\pm$ 0.16 \\
	Beam 2 & 1 & $-$17.93 $\pm$ 0.03 &	...	& 1.01 $\pm$ 0.03 & ... & 20.30 $\pm$ 0.8 & 0.81 $\pm$ 0.05 & 0.10 $\pm$ 0.09	& ... & ... & ... \\ 
	& 2 & $-$3.69 $\pm$ 0.09 & ... & 0.45 $\pm$ 0.09 & ... & 7.18 $\pm$ 0.18 & 0.05 $\pm$ 0.01 & ... & ... & ... & ... \\
	Beam 3 & 1 & $-$16.71 $\pm$ 0.03 & ... & 1.14 $\pm$ 0.03 & ... & 20.30 $\pm$ 0.08 & 0.68 $\pm$ 0.04 & ... & ... & ... & ... \\
	Beam 4 & 1 & $-$17.55 $\pm$ 0.01 & $-$17.69  $\pm$ 0.01 & 0.95 $\pm$ 0.01 & 0.74 $\pm$ 0.01 & 20.30 $\pm$ 0.08 & 11.32 $\pm$ 0.21 & 2.61 $\pm$ 0.10 & ... & ... & ... \\
	& 2 & $-$2.66 $\pm$ 0.02 & ... & 0.36 $\pm$ 0.02 & ... & 5.08 $\pm$ 0.31 & 0.06 $\pm$ 0.01 & ... & ... & ... & ... \\
	Beam 5 & 1 & ... & $-$18.28 $\pm$ 0.02 & ... & 0.73 $\pm$ 0.04 & 20.30 $\pm$ 0.08 & 8.32 $\pm$ 0.14 & 2.36 $\pm$ 0.08 & ... & ... & ... \\
	Beam 6 & 1 & $-$18.35 $\pm$ 0.04\tablenotemark{$\dagger$} & ... & 0.58 $\pm$ 0.04\tablenotemark{$\dagger$} & ... & 20.30 $\pm$ 0.08 & 19.89 $\pm$ 0.27 & 5.23 $\pm$ 0.09 & 4.18 $\pm$ 0.90 & 0.48 $\pm$ 0.14 & 0.65 $\pm$ 0.23 \\ 
	Beam 7 & 1 & $-$17.22 $\pm$ 0.01 & $-$17.52 $\pm$ 0.02 & 1.11 $\pm$ 0.01 & 0.74 $\pm$ 0.02 & 20.30 $\pm$0.08 & 3.63 $\pm$ 0.10 & 1.23 $\pm$ 0.09 & ... & ... & ... \\
	& 2 & $-$1.34 $\pm$ 0.09 & ... & 0.72 $\pm$ 0.09 & ... & 4.76 $\pm$ 0.16 & 0.06 $\pm$ 0.01 & ... & ... & ... & ... 
	\enddata
	\tablenotetext{\dagger}{Determined from the Gaussian fit of the C$^{18}$O spectrum.}
\end{deluxetable*}

\section{Transition between Atomic and Molecular Phase}\label{sec:timescale}

\begin{figure*}[!htb]
	\centering
	\includegraphics[height=0.45\textwidth]{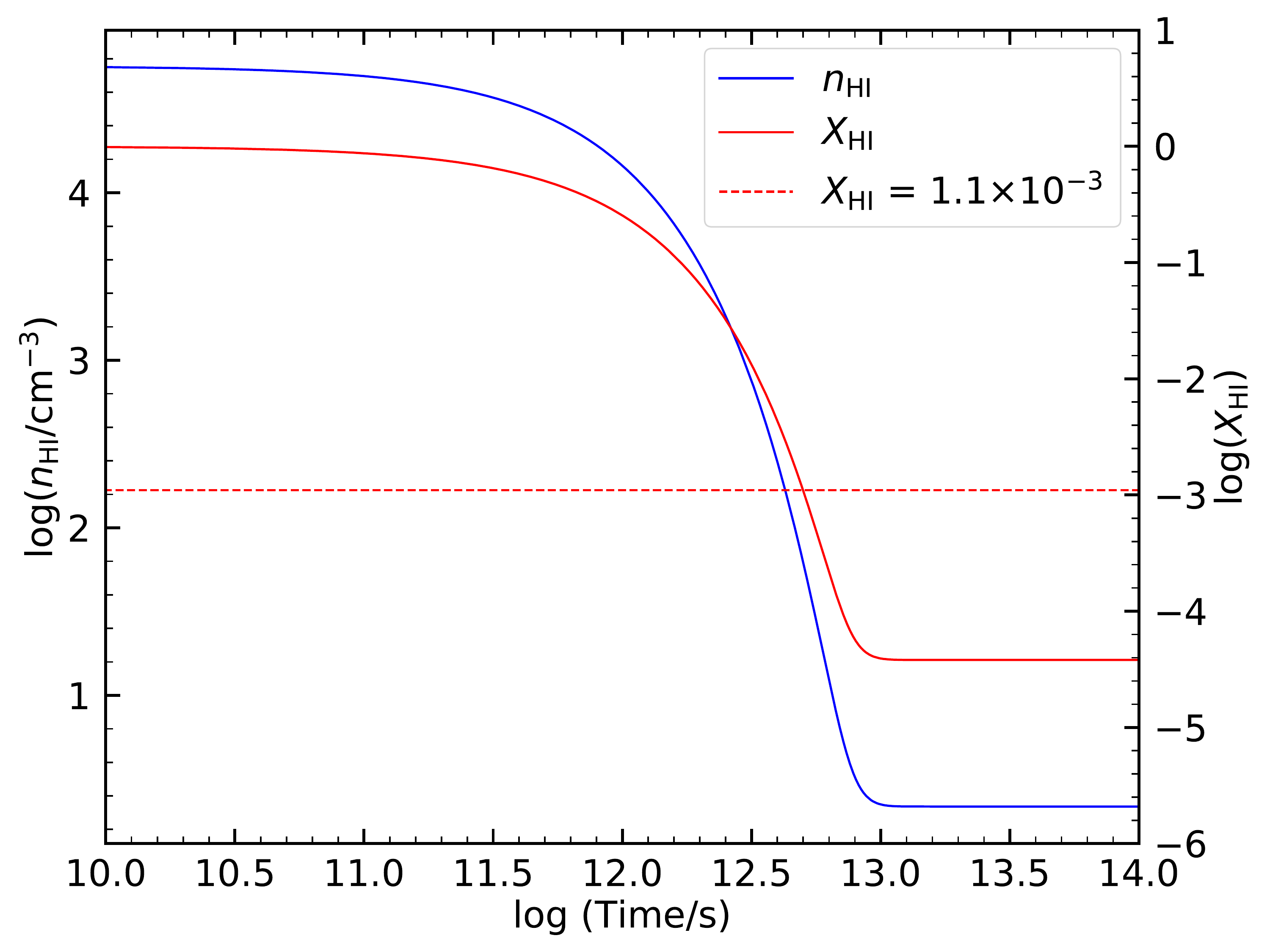}
	\caption{$n_{\mathrm{HI}}$ and $X_{\mathrm{HI}}$ as a function of time. The H~\textsc{i} density in the cloud core drops to its steady state value of $\sim$ 2 cm$^{-3}$ with a time of $\sim$ $4\times10^5$ yr.}
	\label{fig-timescale}
\end{figure*}

We follow the model developed by \cite{Goldsmith+2007} to study the transition between H~\textsc{i} and H$_2$. We only focus on the central constant core, and assume that the central density is $n_0 \sim 5.7\times 10^4$ cm$^{-3}$, line width is 1.14 km s$^{-1}$, radius is $10^{17}$ cm, and $T = 10$ K. The other parameters follow these in \citet{Goldsmith+2007}. The equations (3)--(5) in \citet{Goldsmith+2007} would produce the evolution of the density of H~\textsc{i}, $n_{\mathrm{HI}}$, and H$_2$, $n_{\mathrm{H_2}}$. Figure \ref{fig-timescale} shows $n_{\mathrm{HI}}$ and $X_{\mathrm{HI}}$ as a function of time, where $X_{\mathrm{HI}}$ set to $n_{\mathrm{HI}}/(n_{\mathrm{HI}}+2n_{\mathrm{H_2}})$ assuming that the sizes of cold H~\textsc{i} and CO cloud are similar \citep[see e.g.,][]{Goldsmith+2007}. The time that has elapsed since the material was UV irradiated would be $\sim$ 2 $\times$ 10$^{5}$ yr assuming $X_{\mathrm{HI}} \sim 1.1 \times 10^{-3}$.

\acknowledgments
This work made use of the data from FAST. FAST is a Chinese national mega-science facility, operated by National Astronomical Observatories, Chinese Academy of Sciences. We would like to thank all the staff members of the Delingha 13.7 m radio telescopes for their support making the various molecular line observations. We would like to thank the anonymous referee for the helpful comments and suggestions that helped to improve the paper. This work was sponsored by the Natural Science Foundation of Jiangsu Province (grant No. BK20210999), the Entrepreneurship and Innovation Program of Jiangsu Province, NSFC grants Nos. 11933011 and 11873019, and the Key Laboratory for Radio Astronomy, Chinese Academy of Sciences.

\facility{FAST, PMO 13.7m}


\end{CJK*}
\end{document}